\documentclass[pdflatex,sn-mathphys-num]{sn-jnl}

\usepackage{graphicx}
\usepackage{multirow}
\usepackage{amsmath,amssymb,amsfonts}
\usepackage{amsthm}
\usepackage{mathrsfs}
\usepackage[title]{appendix}
\usepackage{xcolor}
\usepackage{textcomp}
\usepackage{manyfoot}
\usepackage{booktabs}
\usepackage{ragged2e}
\usepackage{algorithm}
\usepackage{algorithmicx}
\usepackage{algpseudocode}
\usepackage{listings}
\theoremstyle{thmstyleone}

\usepackage{soul}

\theoremstyle{thmstyletwo}

\theoremstyle{thmstylethree}

\usepackage{comment}
\usepackage[caption=false]{subfig}
\usepackage{verbatim}
\usepackage[utf8]{inputenc}
\usepackage{gensymb}
\usepackage{natbib}
\usepackage{lineno}
\usepackage{color}
\usepackage{hyperref}
\usepackage{bm}
\usepackage{soul,ulem}
\usepackage[version=3]{mhchem}
\usepackage{tabularx}
\usepackage{caption}

\renewcommand{\thefigure}{\arabic{figure}}

\begin{document}
\def\scititle{Berry Curvature Driven Transport in Silicon-Compatible Altermagnetic $\alpha$-MnTe Thin Films}

\title{\bfseries \boldmath \scititle}
\author[1]{\fnm{Rajib} \sur{Sarkar}}
\author[1]{\fnm{Subhransu Kumar} \sur{Negi}}
\author[2,3]{\fnm{Arindom} \sur{Das}}
\author[2,3]{\fnm{Arijit} \sur{Mandal}}
\author[1]{\fnm{Pankaj} \sur{Bhardwaj}}
\author[1]{\fnm{Sohini} \sur{Guin}}
\author[1]{\fnm{Aryaman} \sur{Das}}
\author[1]{\fnm{Naresh} \sur{Shyaga}}
\author[1]{\fnm{Kartick} \sur{Biswas}}
\author[1]{\fnm{Laxmipriya} \sur{Nanda}}
\author[1]{\fnm{Pavan} \sur{Nukala}}
\author*[2,3]{\fnm{B. R. K. } \sur{Nanda}}\email{nandab@iitm.ac.in}
\author*[1]{\fnm{Dhavala} \sur{Suri}}\email{dsuri@iisc.ac.in}

\affil[1]{\orgdiv{Centre for Nanoscience and Engineering}, \orgname{Indian Institute of Science}, \orgaddress{\street{Bangalore}, \city{Bengaluru}, \postcode{560012}, \state{Karnataka}, \country{India}}}

\affil[2]{\orgdiv{Condensed Matter Theory and Computational Lab}, \orgname{Department of Physics}, \orgaddress{\street{IIT Madras }, \city{Chennai}, \postcode{600036}, \state{Tamil Nadu}, \country{India}}}

\affil[3]{\orgdiv{Center for Atomistic Modelling and Materials Design}, \orgaddress{\street{IIT Madras }, \city{Chennai}, \postcode{600036}, \state{Tamil Nadu}, \country{India}}}

\abstract{Integrating spin-dependent functionality with mainstream semiconductor technology is a central goal of modern spintronics, yet most candidate materials remain incompatible with silicon-based platforms. Here, we report the direct epitaxial integration of $\alpha$-MnTe thin films on Si(111) via molecular beam epitaxy and demonstrate a robust anomalous Hall effect (AHE) in this silicon-compatible altermagnetic system. Despite the absence of net bulk magnetization, the films exhibit a pronounced hysteretic Hall response, providing transport evidence consistent with finite Berry curvature generated by symmetry breaking in the thin-film geometry. High-resolution structural and spectroscopic characterization confirms phase-pure, epitaxial growth with hexagonal NiAs-type symmetry, while magnetotransport measurements reveal correlated hysteresis in both transverse and longitudinal channels with systematic temperature evolution. First-principles calculations reveal substantial uncompensated Berry curvature arising from the spin-split band structure, consistent with altermagnetic symmetry and the origin of the observed Hall response. These results establish MnTe/Si(111) as a silicon-compatible altermagnetic platform and chart a concrete pathway for embedding Berry-phase-driven functionalities into scalable semiconductor device architectures.}

\keywords{Berry curvature, MnTe, Altermagnetism, Anomalous Hall conductivity}

\maketitle

\section{Introduction}\label{sec1}
\maketitle

The realization of spin-dependent transport without net magnetization is a central objective in spintronics, particularly for integration with semiconductor platforms \cite{Zutic_2004,wolf_2001,Dai_2025,Baltz_2018,Din_2024,Kim_2026}. Antiferromagnets offer key advantages including robustness to external perturbations, ultrafast dynamics, and negligible stray fields, yet Berry-curvature-driven transport in such systems remains challenging due to symmetry constraints \cite{Wu_2020,Marti_2014}.

Altermagnetic materials provide a route to overcome this limitation. They exhibit momentum-dependent spin splitting even in the absence of net magnetization. However, this spin splitting alone does not guarantee a net anomalous Hall response, because additional crystal symmetries can enforce cancellation of the Berry curvature over the Brillouin zone. A finite AHC emerges only when the relevant symmetry constraints are lifted \cite{Smejkal_2022,Fender_2025}. This establishes a bridge between compensated magnetic order and topological transport phenomena \cite{Guerrero_2026,Wang_2026,Khatua_2025}. Among candidate systems, manganese telluride (MnTe) is particularly attractive due to its hexagonal NiAs-type structure and high N\'{e}el temperature (310~K)\cite{Kriegner_2017,Li_2022,zhou_2026_expt}. While prior studies have demonstrated altermagnetic signatures in bulk and epitaxial films on lattice-matched substrates \cite{Bey_2025,Chen_2026,Kriegner_2016}, direct integration with silicon remains an outstanding challenge.

Despite this progress, experimental studies on MnTe have so far been largely confined to bulk crystals \cite{Reig_2001,Orlova_2025}, polycrystalline films \cite{Kim_2009}, and  symmetry lowered epitaxial layers grown on lattice-matched III--V substrates~\cite{Bey_2025,Chen_2026,Kriegner_2016}.  Being an insulator with a well-defined gap at the Fermi energy, the bulk altermagnetic MnTe demonstrates vanishing anomalous Hall conductivity and therefore epitaxially grown MnTe films are promising to induce AHC.
In this context, lattice-matched III--V substrates, while enabling high crystalline quality, are fundamentally incompatible with silicon-based device platforms. Achieving epitaxial $\alpha$-MnTe directly on Si(111) is technically challenging \cite{Kim_2009,Lee_2025} due to the substantial lattice mismatch, propensity for interfacial reactions, and phase stability constraints. Addressing this challenge is not merely a materials engineering problem---it is a prerequisite for translating the rich physics of altermagnetism into technologically relevant architectures.

Here we bridge this gap by demonstrating the molecular beam epitaxy (MBE) growth of high-quality epitaxial $\alpha$-MnTe directly on Si(111) and reporting a robust anomalous Hall effect in the resulting films, via a combination of structural characterization, temperature- and field-dependent magnetotransport measurements, and first-principles electronic structure calculations.  In thin films, the $C_{6z}t_{001/2}$ symmetry, which connects the two oppositely spin-polarized sublattices in bulk MnTe, is broken both at the surface and at the film--substrate interface. Consequently, the magnetic point group symmetry of bulk MnTe is modified, intrinsically giving rise to a finite anomalous Hall conductivity (AHC). These results open a concrete and scalable route toward integrating altermagnetic and antiferromagnetic functionalities with semiconductor technology.

\section{Results and Discussions}\label{sec13}

\begin{figure*}[h!]
\centering
\includegraphics[width=\linewidth]{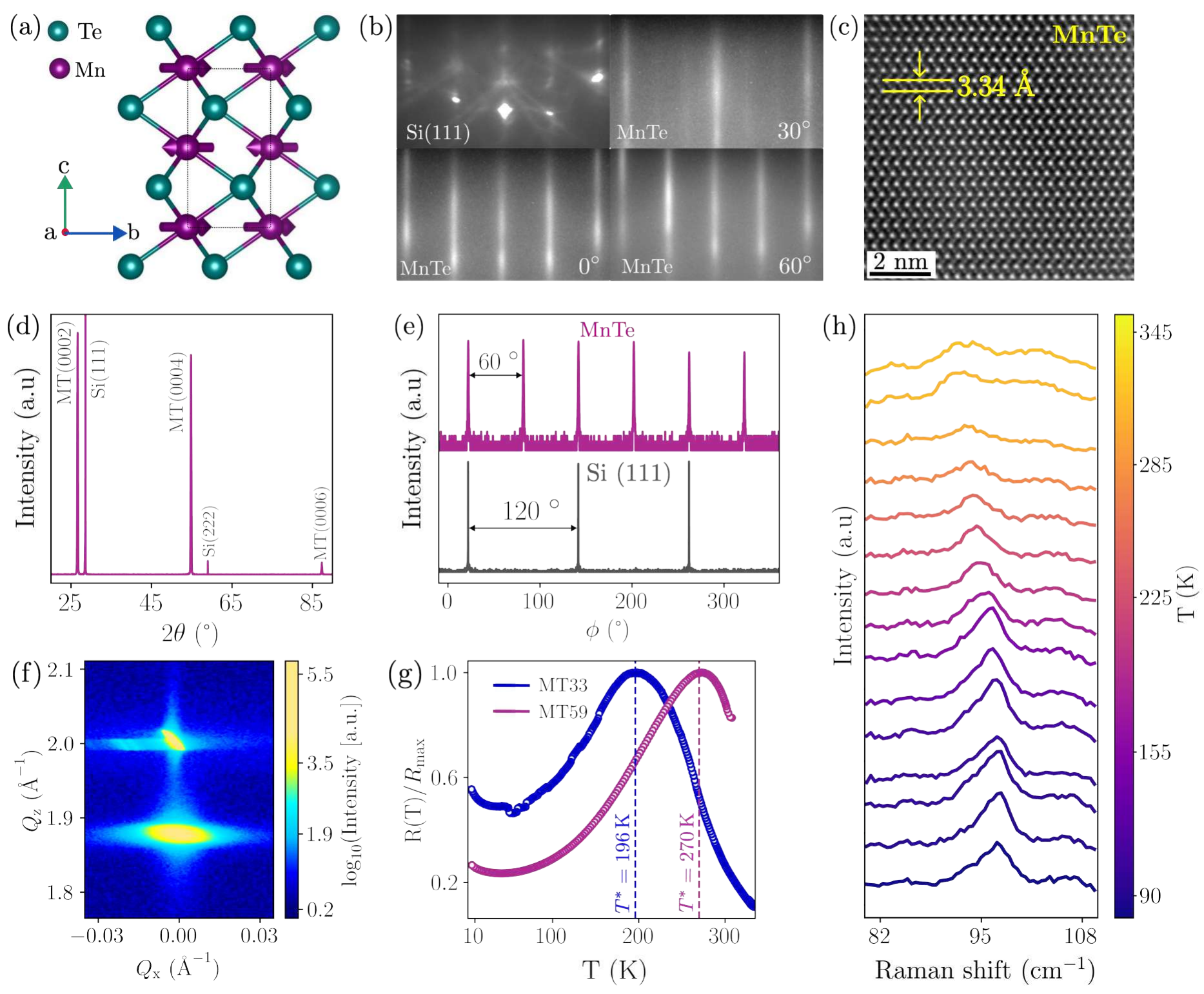}
\caption{
(a) Crystal structure of hexagonal MnTe (NiAs-type) visualized using VESTA (b) Reflection high-energy electron diffraction (RHEED) patterns of the Si(111) substrate and MnTe thin films acquired at different in-plane azimuthal rotations (0$^\circ$, 30$^\circ$, and 60$^\circ$). (c) Cross-sectional transmission electron microscopy (TEM) image of the MnTe film. (d) High-resolution X-ray diffraction (XRD) $\theta$--2$\theta$ scans of MnTe thin films  grown on Si(111), showing a pronounced MnTe (002),(004) and (006) reflection and Si(222) substrate peak.  (e) XRD $\phi$-scans showing the 6-fold rotational symmetry of the MnTe $(10\bar{1}2)$ plane and 3-fold symmetry of the Si(111) substrate. (f) Symmetric reciprocal space mapping (RSM) around the (000L) reflection.   (g) Resistance vs. temperature ($R-T$) curves of two samples MT59 and MT33, samples of different crystalline sample quality. (h) Raman spectra of the MnTe
films displaying the characteristic E$_{2g}$ phonon mode of hexagonal MnTe as a function
of temperature.}
\label{fig:structure_overview}
\end{figure*}

Manganese telluride (MnTe) crystallizes in the hexagonal NiAs-type structure (space group \textit{P6$_3$/mmc})\cite{Takahashi_2025}. In this structure, Te atoms form a hexagonal close-packed (hcp) lattice stacked along the $c$-axis, while Mn atoms occupy all octahedral interstitial sites between adjacent Te layers [Fig.~\ref{fig:structure_overview}(a)]. Each Mn atom is octahedrally coordinated by six Te atoms, and each Te atom is similarly surrounded by six Mn atoms, forming a distorted octahedral environment \cite{Devaraj_2026,Li_2026,Fender_2025}. The crystal can be viewed as alternating layers of Mn and Te stacked along the $c$-axis, while Mn atoms form a hexagonal arrangement within the basal ($ab$) plane. This layered geometry plays a key role in governing the anisotropic electronic and magnetic properties of MnTe. Below the N\'{e}el temperature ($T_{\mathrm{N}} \approx 310$ K), MnTe exhibits antiferromagnetic ordering, where spins are aligned ferromagnetically within the basal plane and antiferromagnetically coupled along the $c$-axis \cite{Dzian_2025,Yamamoto_2025,Dai_2026}.

Reflection high-energy electron diffraction (RHEED) was used to monitor the surface structure \textit{in situ} during growth. The pristine Si(111) substrate exhibits a well-defined streaky RHEED pattern, indicative of a clean, atomically flat surface. Upon initiation of MnTe growth, the pattern evolves [see Appendix~A] into sharp streaks characteristic of two-dimensional, layer-by-layer growth. Azimuthal rotation of the sample about the surface normal reveals a clear rotational symmetry: at an in-plane rotation of $30^\circ$, the RHEED streaks show increased spacing, consistent with probing a distinct crystallographic azimuth, while rotation to $60^\circ$ restores the original streak spacing observed at $0^\circ$ [Fig.~\ref{fig:structure_overview}(b)]. This sixfold periodicity provides direct evidence of the hexagonal in-plane symmetry of the MnTe film, consistent with the expected NiAs-type structure.

From the RHEED streak spacing and X-ray diffraction (XRD) analysis of the MnTe reflections [Fig.~\ref{fig:structure_overview}(d)], we obtain lattice constants $a_{\mathrm{Si(111)}} = 5.44$~\AA\, $a_{\mathrm{MnTe}} = 4.080-4.084$~\AA \ and $c_{\mathrm{MnTe}} = 6.68-6.69$~\AA, which have been used for theoretical calculations.  The error bar in value of the obtained lattice paramaters is less than 1~\%. The MnTe reflections appear exclusively along the $(00l)$ direction, confirming single-phase, highly textured out-of-plane growth, with a full width at half maximum (FWHM) of $\leq 0.2^\circ$, attesting to the high crystalline quality of the film. Tilting the sample to $\chi \approx 42^\circ$ allowed us to probe the epitaxial ordering along an off-axis zone [Fig.~\ref{fig:structure_overview}(e)], revealing hexagonal symmetry in the film and trigonal symmetry in the substrate -- demonstrating that epitaxial ordering extends beyond the in-plane geometry probed by RHEED to this off-axis crystallographic plane. On-axis reciprocal space mapping [Fig.~\ref{fig:structure_overview}(f)] shows the substrate and film reflections aligned along the same reciprocal lattice rod, indicating negligible strain along the $c$-axis. Collectively, these measurements establish the structurally excellent crystalline quality of the MnTe film. Transmission electron microscopy (TEM) confirms this orderly growth [Fig.~\ref{fig:structure_overview}(c)], clearly resolving the Mn and Te sublattices with a lattice constant identical to that obtained from XRD.

\newpage

Having established the high structural quality of the MnTe films, we next examine how this structural fidelity manifests in the electronic and magnetic response of the material. To correlate the electron transport properties with the underlying electronic and magnetic structure, we first examine the temperature-dependent four-point resistance [Fig.~\ref{fig:structure_overview}(g)]. Samples with higher crystalline quality exhibit a correspondingly higher N\'eel temperature than those of lower crystallinity, indicating that the antiferromagnetic ordering is highly sensitive to structural quality, consistent with theoretical predictions \cite{Junior_2023}. This sensitivity of the magnetic transition to sample quality motivates a closer, microscopic probe of the interplay between lattice and magnetic degrees of freedom, for which Raman spectroscopy offers a direct and complementary signature. Interestingly, the Raman mode associated with the $E_{2g}$ phonon at $\approx 98$~cm$^{-1}$ displays a pronounced temperature dependence [Fig.~\ref{fig:structure_overview}(h)], emerging  only below $230$~K and decaying above this temperature. This behavior is attributed to spin-phonon coupling, wherein the onset of magnetic ordering modifies the phonon vibrational characteristics -- a phenomenon well documented in several magnetic systems, including MnTe \cite{Zhang_2020,shao_2026}. Taken together, the transport and Raman signatures confirm that the magnetic and structural degrees of freedom in these films are strongly and consistently coupled, underscoring that the high crystalline quality established above translates directly into well-defined magnetic ordering.

The synthesis of MnTe on Si(111) via this approach represents a significant milestone, as it establishes a route to integrating an altermagnetic material directly with the technologically dominant semiconductor platform. This growth is not enabled by any symmetry reduction, but is instead achieved purely through domain-matching epitaxy. Consequently, the electronic properties are expected to be comparable to those of MnTe synthesized on InP substrates, while offering the added compatibility with existing Si-based device architectures.

To understand the electronic structure and the results in MnTe film, DFT+U calculations are carried out using plane-wave based projector augmented wave methods with lattice parameters derived from RHEED and XRD experiments \cite{Blochl1994-wp, PhysRevB.59.1758}, and implemented in Vienna \textit{Ab initio} Simulation Package (VASP). For the calculations, the PBE-GGA exchange-correlation functional \cite{ernzerhof1998generalized} is used, and a $\Gamma$-centered $12 \times 12 \times 1$ k-mesh is employed for the Brillouin zone (BZ). The energy cutoff to construct the plane-wave basis set is $400$ eV. To account for the strong correlation effect, we set the effective Hubbard parameter $U_{\text{eff}} = U - J$ to $3$ eV \cite{Devaraj_2026, Nirmal_2024, PhysRevLett.130.036702} within the rotationally invariant framework proposed by Dudarev et al.\ \cite{dudarev}. A symmetry analysis \cite{PhysRevB.111.184407, Das_2026} implies the breakdown of the $\tau C_{6z}$ or $\tau S_{6z}$ in bulk MnTe, allowing the existence of AHC with spin quantization axis along $y$ and $z$ direction. However, for the case of MnTe films, as these symmetries vanish, the spin quantization along $z$ axis can give finite AHC at the Fermi level (see Appendix M, Fig. M19). Both the spin quantization axes are considered in this study. For the calculation of Berry curvature, and thereby the AHC, maximally localized Wannier functions are obtained with the aid of the Wannier90 code \cite{MOSTOFI2008685}. For this purpose, a dense k-mesh of $300 \times 300 \times 1$ is used with an adaptive k-mesh of $5 \times 5 \times 1$ for a threshold value of $100\,\text{bohr}^2$. 

\begin{figure}[ht!]
    \centering
    \includegraphics[width=0.65\linewidth]{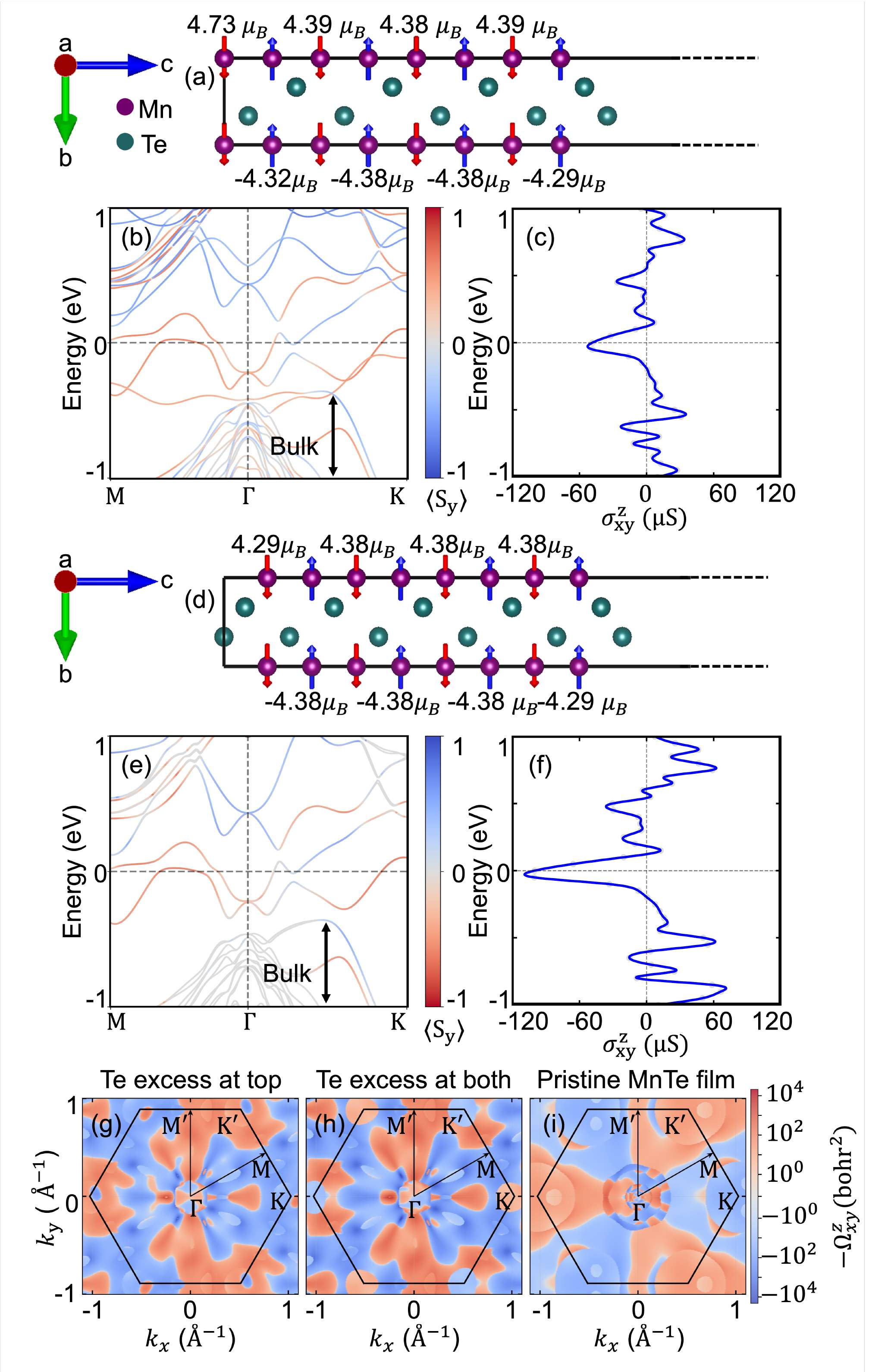}
    \caption{(a) Stacking arrangements of config-I, where extra Te capping at the top with Mn layer exposed to the substrate, and the corresponding spin ($\langle S_y \rangle$) projected DFT+SOC bandstructure (b) and AHC (c). (d) Stacking arrangements of config-II, where extra Te capping is both at the top and bottom, and the corresponding spin projected DFT+SOC bandstructure (e) and AHC (f). The magnetic moment (in $\mu_B$) at each Mn site for both the configurations are indicated. (g--i) The Berry curvature distribution ($\Omega^{z}_{xy}$) over the BZ for config-I, config-II, and config-III (pristine MnTe film; see Fig.~M18 in Appendix). As can be implied from Fig.~\ref{fig:DOS_top_both_Te}, the surface bands are occupying the Fermi level, while the interior MnTe layers maintain the antiferromagnetic insulating behavior. The AHC with spin quantization along the $z$ axis is provided in Fig.~M19 in Appendix M.}
    \label{fig:top_both_Te_Im2}
\end{figure}

We investigate a pristine MnTe film with a Te-terminated top layer and an
Mn-terminated bottom layer (config-III), with the corresponding
results presented in Appendix~M, Fig. M18. Similar to bulk MnTe, this configuration
exhibits positive and negative contributions to $\Omega_{xy}^{z}$ that are
nearly equally compensated, except in the vicinity of the high-symmetry
point $\Gamma$, resulting in a weak net AHC. At this stage, however, an
important experimental consideration must be addressed: MnTe films are
typically synthesized under Te-rich growth conditions. It is therefore
essential to model a slab configuration that reflects this excess-Te
environment rather than the pristine, stoichiometric case. Accounting for
this, we design two additional MnTe slab configurations, denoted
config-I and config-II, as shown in Fig.~\ref{fig:top_both_Te_Im2}. Config-I is a highly certain possibility given the experimentally done Te capping. Note that the excess-Te condition is modeled in this manner; however, this does not imply that the sample separates into distinct surface and bulk states.

In config-I, the Te termination, along with an additional capping layer, is added on the top, while the bottom layer, which is expected to be on the substrate, is Mn-terminated. In config-II, both top and bottom are Te-terminated with additional Te capping layers. A total of 8 MnTe layers are considered, which is sufficient to capture both the surface and bulk (interior) electronic structures. To account for the substrate effect, the experimentally obtained in-plane lattice parameters of the MnTe films are $a = b = 4.08$~\AA. The inter-layer Mn--Mn separation is taken to be that of the bulk, i.e., $3.35$~\AA, and, to avoid the spurious periodic interactions, a vacuum of $15$~\AA\ is placed on the constructed MnTe slabs. Since the growth substrate is an insulator, Si(111), for all practical purposes can be treated as a vacuum and hence, is not considered in this study to reduce the computational expenses. The total energy calculations (see Table~M4 in Appendix M) suggest that the MnTe films retain the AFM configurations, which is in agreement with the experimental observation made in this work. An energy comparison of the FM and A-type AFM (altermagnetic) reveals that like bulk, the latter stabilize the magnetic configuration in films.

\begin{figure}[ht!]
    \centering
    \includegraphics[width=0.8\linewidth]{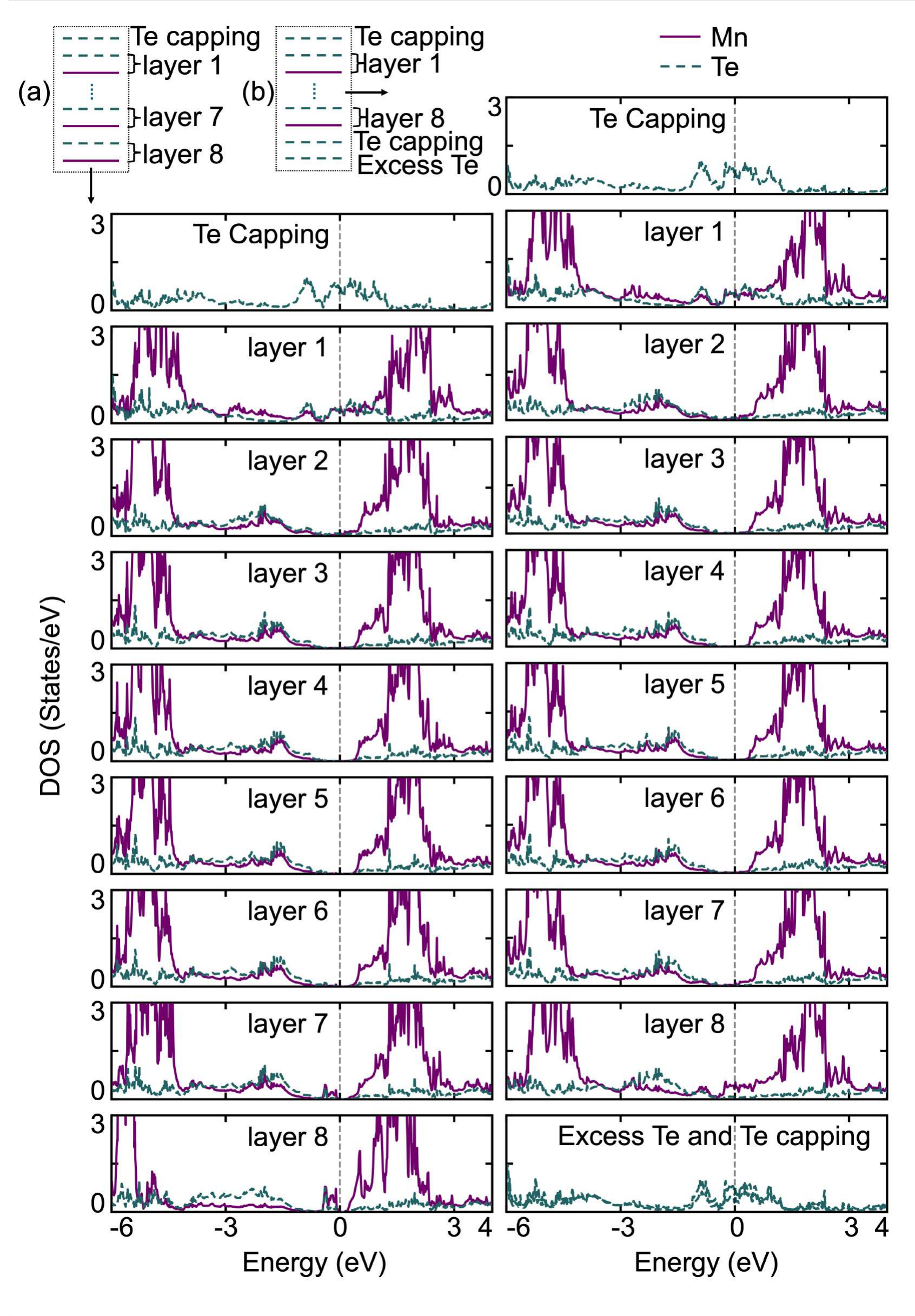}
    \caption{Layer projected DOS for config-I (left) and config-II (right). Except the capping layer, each layer consists of one Mn and the adjacent Te atomic layers as schematically indicated. Beside the capping layers, the states of the surface layers occupy the Fermi level.}
    \label{fig:DOS_top_both_Te}
\end{figure}

The DFT+U+SOC AFM bandstructure of config-I and -II are shown in Fig.~\ref{fig:top_both_Te_Im2}(b) and (e). From the figure we gather that, a spectrum of bands crosses the Fermi level validating the experimentally observed metallicity. When compared with the insulating bulk band structure (see Appendix M, Fig.~M16), the bands emerging from the interior layers could easily be identified and denoted as bulk. For further substantiation, the layer projected Mn and Te resolved density of states (DOS) are shown in Fig.~\ref{fig:DOS_top_both_Te}. For both the configurations, the states occupying the Fermi level are formed by the capping Te as well as top and bottom Mn/Te layers. The rest of the layers replicate the bulk insulating electronic structure. For these interior layers, due to strong correlation, the occupied spin majority and unoccupied spin minority states are far apart and lie away from the Fermi level. The local spin moment of each Mn layer, as indicated in Fig.~\ref{fig:top_both_Te_Im2}, suggests that the interior layers follow the ideal bulk altermagnetic configurations. However, the top and bottom Mn bilayers exhibit uncompensated magnetization. From the symmetry point of view it can be explained as follows. In the bulk MnTe, the two opposite spin sublattices in this A-type altermagnetic system are connected through $C_{6z}t_{0\,0\,1/2}$ which ensures a compensated magnetization as one observes in an ideal antiferromagnet. The absence of this symmetry connection between the top/bottom Mn bilayers, even though there is an antiparallel spin orientation within the bilayer, a resulting magnetization emerge. Interestingly, in the case of config-II, the magnetization of top bilayer is equal and opposite to that of the bottom bilayer to make the film as a whole completely spin compensated. However, for config-I, we notice residual magnetization in the film which is largely contributed by the bottom Mn layer.

The AHC is estimated by using the Kubo formula involving the Berry curvature $\Omega_{n}^{z}(\mathbf{k})$ \cite{PhysRevLett.49.405,doi:10.1143/JPSJ.12.570},

\begin{equation}
 \sigma_{xy}^{z} = \frac{e^2}{\hbar} \sum_{n} \int_{\mathrm{BZ}} f(\epsilon_{n}(\mathbf{k}))\, \Omega_{n}^{z}(\mathbf{k})\, \frac{d^{2}\mathbf{k}}{(2\pi)^2},
 \label{sigma_eqn}
\end{equation}

\begin{equation}
    \Omega_n^z (\boldsymbol{k}) = -2\hbar^2\, \mathrm{Im} \sum_{m \neq n} \frac{\langle \psi_{m\boldsymbol{k}} | v_x |\psi_{n\boldsymbol{k}} \rangle \langle \psi_{n\boldsymbol{k}} | v_y |\psi_{m\boldsymbol{k}} \rangle}{(\epsilon_{n\boldsymbol{k}} - \epsilon_{m\boldsymbol{k}})^2}.
    \label{berry_curv}
\end{equation}

Here, $e$, $\hbar$, $n$, and $f(\epsilon_{n}(\mathbf{k}))$ represent the electron charge, reduced Planck's constant, band index, and the Fermi-Dirac distribution function, respectively. $v_{x(y)} = (\hbar^{-1})\,\partial\mathcal{H}/\partial k_{x(y)}$ is the velocity operator, and $\epsilon_{n\boldsymbol{k}}$ is the energy eigenvalue of the $n$-th band. The Berry curvature is mapped across the BZ and its shown in Fig.~M23 of Appendix, for all three configurations. The finite values of integration leads to AHC which is plotted in Fig.~\ref{fig:top_both_Te_Im2}(c) and (f). Contrasting to the excess Te film, the pristine MnTe film shows weak AHC at the Fermi level despite having surface states [see Appendix M, Fig.~M18]. This can be understood from the distribution of the Berry curvature ($\Omega_{xy}^z$). All the three configurations demonstrate the expected three-fold symmetry. For config-I and -II, the distribution of negative $\Omega_{xy}^z$ (blue) is much larger than the positive $\Omega_{xy}^z$ (red), giving rise to the large finite AHC. For config-III, the distribution of negative and positive $\Omega_{xy}^z$ is nearly equal to form a vanishing AHC ($\approx 0.6\,\mu$S). A detailed analysis of the bandstructure along the $(-K)$--$\Gamma$--$K$ [as shown in M20 of Appendix M] infers that, for config-II there is a Kramers spin degeneracy of the bands in the absence of SOC as in the case of bulk. The lifting of Kramers spin degeneracy (LKSD) with the SOC is responsible for the finite AHC. The LKSD is attributed to the altermagnetic characteristic of $\alpha$-MnTe \cite{Krempasky_2024,wang_2026_InP_MnTe,zhao_2026_theory}. For config-I and -III, the LKSD occurs both due to SOC and TRS breaking. We find that if the spin quantization is taken along the $z$ axis, the system still exhibits finite AHC.

\begin{figure*}[t]
\centering
\includegraphics[width=\textwidth]{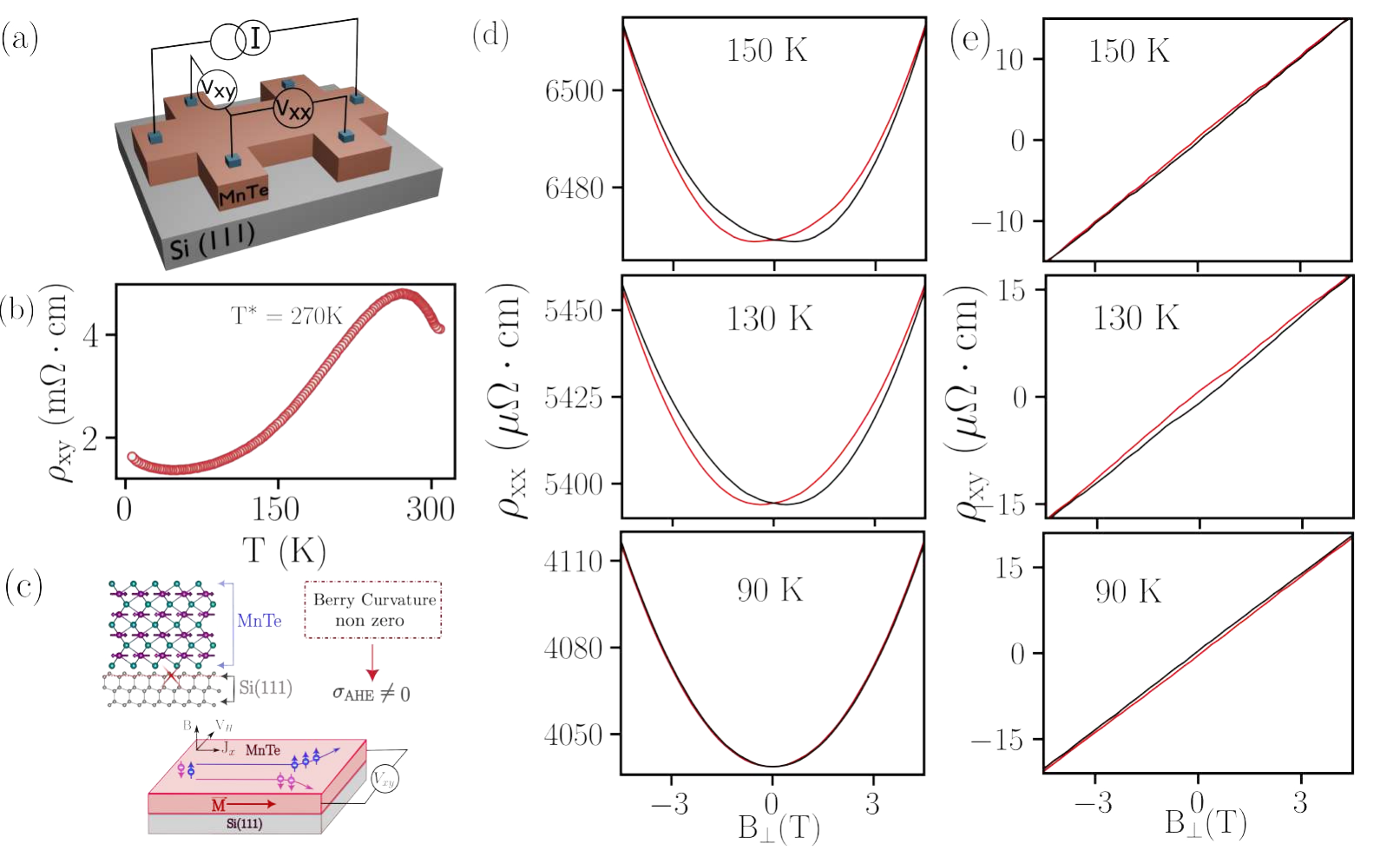}
\caption{(a) Schematic of the Hall bar device fabricated from epitaxial MnTe film illustrating the standard four-probe geometry used for simultaneous longitudinal ($V_{xx}$) and transverse ($V_{xy}$) voltage measurements. (b) Resistance plotted against temperature. (c) Illustration showing the origin of non-zero Berry curvature and hence a finite anomalous Hall effect in MnTe on Si(111).
(d) Longitudinal resistivity $\rho_{xx}$ as a function of $B_\perp$ at different temperatures as mentioned in the legend. (e) Hall resistivity $\rho_{xy}$ as a function of $B_\perp$ measured at different temperatures as mentioned in the legend. These measurements are performed on a sample identical to the sample whose structural characteristics are shown in Fig. 1}
\label{fig:transport}
\end{figure*}

Upon simulating this excess-Te condition, we find that the anomalous Hall
conductivity is non-zero, in marked contrast to the compensated response
obtained for the pristine, stoichiometric configuration. This result
suggests that Te-rich growth conditions, which are inherent to the
experimental synthesis of MnTe films, can intrinsically break the
compensation between positive and negative $\Omega_{xy}^{z}$ contributions
and give rise to a finite AHC. 

Motivated by this finding, we further examine the transport measurements on our experimentally grown MnTe films
to determine whether a similar anomalous Hall signature is observed. To do so, we first scribe a Hall bar from MnTe film, via LASER cutting [Fig.~\ref{fig:transport}(a) shows the illustration]. This ensures that no lithographic process damages the sample surface. The resistance versus temperature curve of MnTe is shown in Fig.~\ref{fig:transport}(b). It exhibits a distinct hump around $\approx$270~K indicating maximal spin scattering at this temperature. Below these temperatures, anti-ferromagnetic ordering reduces the spin scattering and hence the drop in resistance \cite{Kriegner_2016,wu_2025,Magnin_2012}. While rotational symmetry is broken in the sample due to local spin ordering in the crystal structure, inversion symmetry is broken due to the interface between substrate and the epitaxial MnTe film. The resultant band structure exhibits a non-zero Berry curvature giving rise to finite AHC as illustrated in Fig.~\ref{fig:transport}(c). To investigate into this hypothesis we perform magnetoresistance and Hall resistance measurements with magnetic field perpendicular to the sample plane.

Fig.~\ref{fig:transport}(d--e) reveal transport signatures consistent with a Berry-curvature-driven response. A correlated hysteresis in both longitudinal and transverse channels indicates a common origin in magnetic domain dynamics. The longitudinal resistance exhibits maximal hysteresis in the temperature around $\approx$ 150~K, consistent with anisotropic magnetoresistance governed by the N\'{e}el vector orientation, while weak domain pinning leads to pronounced history dependence despite zero net magnetization [see Appendix L]. The anti-symmetrized Hall resistance also shows clear hysteresis, evidencing an anomalous Hall contribution beyond the ordinary response.

To further elucidate the origin of the anomalous Hall effect, we analyze the scaling between $\sigma_{xy}$ and $\sigma_{xx}$ (see Appendix G): unlike extrinsic mechanisms, which produce characteristic scaling relations, the two quantities here evolve independently, consistent with an intrinsic origin \cite{Nagaosa_2010,Siddiquee2023}. Magnetization measurements (see Appendix L) show an extremely weak magnetic signature, possibly arising from canting  due to interfacial strain (other possible reasons, are discussed on the SI), thereby making a dominant conventional ferromagnetic contribution to the observed AHE unlikely. We have also investigated samples with thicknesses of 10 nm, 20 nm, 50 nm and 200 nm (see Appendix F). Across this series, the anomalous Hall resistivity increases with thickness before saturating at larger thickness (Appendix G), while a nonzero anomalous Hall response is present at every thickness studied, including the thinnest (10 nm) film. We interpret this trend as reflecting two additive contributions to the total AHC: an interfacial/strain-induced contribution, expected to be comparable across samples regardless of thickness, and a surface/bulk altermagnetic contribution that scales with the thickness of the active magnetic layer before saturating once the film thickness exceeds the relevant strain-relaxation and electron-penetration depths. The weaker hysteresis observed in the thinnest (10 nm) film is therefore attributed to its smaller altermagnetic contribution relative to the (thickness-independent) interfacial term, rather than to a suppression of hysteresis in the single-domain limit; a single-domain film is expected to switch coherently under an applied field and would not, by itself, produce reduced hysteresis. This picture is consistent with the intrinsic, Berry-curvature-driven mechanism described above, and with recent reports attributing the AHE in altermagnetic MnTe to combined surface and bulk contributions whose relative weight evolves systematically with thickness. In bulk MnTe, symmetry operations enforce cancellation of Berry curvature; however, in thin films, symmetry breaking---arising from interfacial inversion asymmetry, epitaxial strain lifts this constraint and enables a finite AHC. Field-driven reconfiguration of antiferromagnetic domains modifies both carrier scattering and the Berry curvature distribution in momentum space. Taken together, these observations support an intrinsic mechanism in which symmetry lowering enables Berry-curvature-driven transport in the altermagnetic state.

Finally, we discuss the carrier mobility of the MnTe samples, which we find to be notably high compared to conventional semiconductors [SI Section~H]. The thinner MnTe sample exhibits a mobility of approximately 120~cm$^{2}$/V$\cdot$s, remarkably high even relative to bulk MnTe. This enhanced mobility in the thinner sample indicates that surface states, owing to their metallic character, govern the sample's conductivity---consistent with our DFT results, which predict a metallic surface state for the MnTe/Te layer. In thicker samples, surface--bulk hybridization diminishes this surface-dominated behavior, resulting in reduced conductivity and mobility. Interestingly, we find that the AHE signal increases with increase in thickness (refer to SI), which clarifies the genesis of transverse conductivity, that the contribution is due to both the surface and bulk. It is not possible to eliminate either of them completely.  The combined observation of an anomalous Hall effect without a corresponding bulk net magnetization, correlated hysteresis in the longitudinal and transverse channels, weak scaling between \(\sigma_{xy}\) and \(\sigma_{xx}\), and a finite calculated AHC following thin-film symmetry lowering supports a predominantly intrinsic Berry-curvature contribution to the observed Hall response.

\section{Materials and Methods}\label{sec13}

The films were synthesized using molecular beam epitaxy (MBE) under ultra-high vacuum conditions with a base pressure of approximately \(1 \times 10^{-10}\) mbar. High-purity Mn and Te were evaporated from standard effusion cells. Si(111) substrates were initially cleaned using a 5\% HF solution for 1 minute to remove the native oxide layer and obtain a hydrogen-terminated surface. The substrates were subsequently introduced into the MBE chamber and annealed at \(700^\circ\mathrm{C}\) for 30 minutes to promote surface reconstruction and desorption of residual contaminants. Following this, the substrate temperature was gradually reduced to \(420^\circ\mathrm{C}\) to initiate the growth. The Mn:Te flux ratio was maintained at 1:5 to promote stoichiometric growth under Te-rich conditions. All films were grown using a single-step growth process at this substrate temperature.

\section{Conclusion}\label{sec13}

In conclusion, we demonstrate synthesis of MnTe thin films directly integrated on Si(111), establishing a silicon-compatible platform for altermagnetic functionality.  Our theoretical studies suggest that the presence of excess Te enhances AHC almost by two orders of magnitude.  The emergence of a finite, hysteretic Hall response in a compensated antiferromagnet highlights the critical role of symmetry breaking in enabling Berry-curvature-driven transport, as supported from our theoretical calculations. These results provide a viable pathway for integrating altermagnetic materials into semiconductor-based spintronics architectures.

\backmatter

\bmhead{Supplementary information}

\noindent \justifying Additional supporting information can be found online in the Supporting Information section.

\bmhead{Acknowledgements}

DS thanks IISc start-up grant, Ministry of Electronics and Technology, Indian Space Research Organization for funding. Authors duly acknowledge funding from INOXCVA and INOX Airproducts for funding via CSR grants. Authors are grateful to micro and nano characterization facility, CeNSE and national nanofabrication facility, CeNSE for facilities usage. PB thanks Anusandhan National Research Foundation for National Postdoctoral Fellowship. LPN thanks DST-National Postdoctoral Fellowship in Nano Science and Technology for fellowship.

\section*{Declarations}

\section*{Conflict of interest}\quad The authors declare no competing
interests.

\section*{Data availability}\quad The data that support the findings of
this study are available from the corresponding authors upon reasonable
request.
\section*{Funding} This work was supported Ministry of Electronics and Information Technology, Govt. of India, INOX Airproducts, INOXCVA and Infosys foundation. Postdoctoral fellowships were funded by Department of Science and Technology, Govt. of India via Anusandhan National Research Foundation and Nano PDF program. 

\section*{Author Contributions} Idea was conceived by RS and DS. RS, SKN grew the samples in MBE. RS and SKN performed electronic transport measurements. AD, AM and BRKN performed density functional theory calculations. PB, SG, AD, NS and LN supported the characterization experiments. DS and BRKN wrote the manuscript with inputs from all authors. DS supervised the entire project. 

\section*{Competing interests} The authors declare that they have no competing interests.

\section*{Data and materials availability} All data needed to evaluate the conclusions in the paper are present in the paper and/or the Supplementary Materials.

\bibliography{references.bib}

@article{Zutic_2004,
  title = {Spintronics: Fundamentals and applications},
  author = { Igor and Fabian, Jaroslav and Das Sarma, S.},
  journal = {Rev. Mod. Phys.},
  volume = {76},
  issue = {2},
  pages = {323--410},
  numpages = {0},
  year = {2004},
  month = {Apr},
  publisher = {American Physical Society}
}

@article{Zhou2025,
  title     = "In-plane Hall effect in a {RuO2} single crystal",
  author    = "Zhou, Xuebo and Yao, Yugui and Li, Zheng and Cao, Jin and Wu,
               Wei and Luo, Jianlin",
  journal   = "J. Phys. Chem. Lett.",
  publisher = "American Chemical Society (ACS)",
  volume    =  16,
  number    =  31,
  pages     = "7883--7888",
  month     =  aug,
  year      =  2025,
  language  = "en"
}

@ARTICLE{Wang2025,
  title     = "Unveiling an in-plane Hall effect in rutile {RuO2} films",
  author    = "Wang, Meng and Zhang, Jianbing and Tian, Di and Yu, Pu and
               Kagawa, Fumitaka",
  journal   = "Commun. Phys.",
  publisher = "Springer Science and Business Media LLC",
  volume    =  8,
  number    =  1,
  pages     = "28",
  month     =  jan,
  year      =  2025,
  copyright = "https://creativecommons.org/licenses/by-nc-nd/4.0",
  language  = "en"
}

@article{wolf_2001,
  author  = {Wolf, Stuart A. and Awschalom, David D. and Buhrman, Robert A. and Daughton, James M. and von Moln{\'a}r, Stephen and Roukes, Michael L. and Chtchelkanova, Alla Y. and Treger, David M.},
  title   = {Spintronics: a spin-based electronics vision for the future},
  journal = {Science},
  year    = {2001},
  volume  = {294},
  number  = {5546},
  pages   = {1488--1495},
  pmid    = {11711666}
}

@article{Baltz_2018,
  title = {Antiferromagnetic spintronics},
  author = {Baltz, V. and Manchon, A. and Tsoi, M. and Moriyama, T. and Ono, T. and Tserkovnyak, Y.},
  journal = {Rev. Mod. Phys.},
  volume = {90},
  issue = {1},
  pages = {015005},
  numpages = {57},
  year = {2018},
  month = {Feb},
  publisher = {American Physical Society}
}

@article{Dai_2025,
author = {Dai, Bingqian and Cheng, Yang and Qu, Tao and Huang, Puyang and Li, Yaochen and Wang, Tianyi and Huang, Hanshen and Shu, Qingyuan and Wang, Kang L.},
title = {Antiferromagnetic Materials Exhibiting Unconventional Properties},
journal = {Adv. Funct. Mater.},
volume = {35},
number = {52},
pages = {e08282},
year = {2025}
}

@article{Din_2024,
  author  = {Dal Din, A. and Amin, O. J. and Wadley, P. and Edmonds, K. W.},
  title   = {Antiferromagnetic spintronics and beyond},
  journal = {npj Spintronics},
  year    = {2024},
  volume  = {2},
  number  = {1},
  pages   = {25},
  issn    = {2948-2119}
}

@article{Kim_2026,
author = {Kim, Yun-Ho and Kim, Gil-Sung and Choi, Jae Won and Cho, Jung-Min and Lee, Won-Yong and Yoon, Hongkee and Choi, Yeonho and Fields, Shelby and Bennett, Steven and Zebarjadi, Mona and Yoon, Young-Gui and Lee, Sang-Kwon},
title = {Large Anomalous Hall Effect, Non-Vanishing Berry Curvature in (110) FeRh Antiferromagnet Films via Interface Strain},
journal = {Adv. Sci..},
volume = {13},
number = {19},
pages = {e18999},
year = {2026}
}

@article{Smejkal_2022,
   title={Beyond Conventional Ferromagnetism and Antiferromagnetism: A Phase with Nonrelativistic Spin and Crystal Rotation Symmetry},
   volume={12},
   ISSN={2160-3308},
   
   journal={Phys. Rev. X},
   publisher={American Physical Society (APS)},
   author={Šmejkal, Libor and Sinova, Jairo and Jungwirth, Tomas},
   year={2022},
   month=sept }

@article{Betancourt_2023,
  title     = {Spontaneous Anomalous Hall Effect Arising from an Unconventional Compensated Magnetic Phase in a Semiconductor},
  author    = {Gonzalez Betancourt, R. D. and Zuba{\v{c}}, J. and Gonzalez-Hernandez, R. and Geishendorf, K. and {\v{S}}ob{\'{a}}{\v{n}}, Z. and Springholz, G. and Olejn{\'{\i}}k, K. and {\v{S}}mejkal, L. and Sinova, J. and Jungwirth, T. and Goennenwein, S. T. B. and Thomas, A. and Reichlov{\'{a}}, H. and {\v{Z}}elezn{\'{y}}, J. and Kriegner, D.},
  journal   = {Phys. Rev. Lett.},
  volume    = {130},
  number    = {3},
  pages     = {036702},
  year      = {2023},
  publisher = {American Physical Society}
}

@article{Krempasky_2024,
  author  = {J. Krempask{\'y} and L. {\v{S}}mejkal and S. W. D'Souza and M. Hajlaoui and G. Springholz and K. Uhl{\'\i}{\v{r}}ov{\'a} and F. Alarab and P. C. Constantinou and V. Strocov and D. Usanov and W. R. Pudelko and R. Gonz{\'a}lez-Hern{\'a}ndez and A. Birk Hellenes and Z. Jansa and H. Reichlov{\'a} and Z. {\v{S}}ob{\'a}{\v{n}} and R. D. Gonzalez Betancourt and P. Wadley and J. Sinova and D. Kriegner and J. Min{\'a}r and J. H. Dil and T. Jungwirth},
  title   = {Altermagnetic lifting of Kramers spin degeneracy},
  journal = {Nature},
  year    = {2024},
  volume  = {626},
  number  = {7999},
  pages   = {517--522},
  issn    = {1476-4687}
}

@misc{zhou_2026_expt,
      title={Surface-State-Driven Anomalous Hall Effect in Altermagnetic MnTe Films}, 
      author={Ling-Jie Zhou and Zi-Jie Yan and Hongtao Rong and Yufei Zhao and Pu Xiao and Lok-Kan Lai and Zhiyuan Xi and Ke Wang and Tibendra Adhikari and Ganesh P. Tiwari and Zhong Lin and Pascal Manue and Fabio Orlandi and Dmitry Khalyavin and Alexander J. Grutter and Chao-Xing Liu and Binghai Yan and Cui-Zu Chang},
      year={2026},
      archivePrefix={arXiv},
      primaryClass={cond-mat.mtrl-sci}
}

@article{Orlova_2025,
  title = {Magnetization symmetry for the altermagnetic candidate MnTe},
  author = {Orlova, N. N. and Esin, V. D. and Timonina, A. V. and Kolesnikov, N. N. and Deviatov, E. V.},
  journal = {Phys. Rev. B},
  volume = {111},
  issue = {22},
  pages = {224414},
  numpages = {8},
  year = {2025},
  month = {Jun},
  publisher = {American Physical Society}
}

@article{Yamamoto_2025,
  title = {Altermagnetic nanotextures revealed in bulk $\mathrm{Mn}\mathrm{Te}$},
  author = {Yamamoto, Rikako and Turnbull, Luke Alexander and Schmidt, Marcus and Corsaletti Filho, Jos'e Claudio and Binger, Hayden Jeffrey and Di Pietro Mart\'{\i}nez, Marisel and Weigand, Markus and Finizio, Simone and Prots, Yurii and Ferguson, George Matthew and Vool, Uri and Wintz, Sebastian and Donnelly, Claire},
  journal = {Phys. Rev. Appl.},
  volume = {24},
  issue = {3},
  pages = {034037},
  numpages = {11},
  year = {2025},
  month = {Sep},
  publisher = {American Physical Society}
}

@article{Reig_2001,
title = {Growth and characterisation of MnTe crystals},
journal = {J. Cryst. Growth},
volume = {223},
number = {3},
pages = {349-356},
year = {2001},
issn = {0022-0248},
author = {C. Reig and V. Muñoz and C. Gómez and Ch. Ferrer and A. Segura}
}

@ARTICLE{Kim_2009,
  author={Kim, Woochul and Park, Il Jin and Kim, Hyung Joon and Lee, Wooyoung and Kim, Sam Jin and Kim, Chul Sung},
  journal={IEEE Trans. Magn.}, 
  title={Room-Temperature Ferromagnetic Property in MnTe Semiconductor Thin Film Grown by Molecular Beam Epitaxy}, 
  year={2009},
  volume={45},
  number={6},
  pages={2424-2427}}

@article{Bey_2025,
  title = {Interface, bulk and surface structure of heteroepitaxial altermagnetic \ensuremath{\alpha}-MnTe films grown on GaAs(111)},
  author = {Bey, Sara and Zhukovskyi, Maksym and Orlova, Tatyana and Fields, Shelby and Lauter, Valeria and Ambaye, Haile and Ievlev, Anton and Bennett, Steven P. and Liu, Xinyu and Assaf, Badih A.},
  journal = {Phys. Rev. Mater.},
  volume = {9},
  issue = {7},
  pages = {074404},
  numpages = {8},
  year = {2025},
  month = {Jul},
  publisher = {American Physical Society},
  
}

@article{Chen_2026,
  author  = {Chen, An-Hsi and Raghuvanshi, Parul R. and Cook, Jacob and Chilcote, Michael and Lapano, Jason and Mazza, Alessandro R. and Lu, Qiangsheng and Kim, Sangsoo and Wu, Yueh-Chun and Ward, T. Zac and Lawrie, Benjamin J. and Bian, Guang and Burns, James and Poplawsky, Jonathan D. and Han, Myung-Geun and Zhu, Yimei and Lindsay, Lucas and Miao, Hu and Gai, Zheng and Moore, Robert G. and Eres, Gyula and Cooper, Valentino R. and Brahlek, Matthew},
  title   = {Programmable Phase Selection between Altermagnetic and Noncentrosymmetric Polymorphs of MnTe on InP via Molecular Beam Epitaxy},
  journal = {ACS Appl. Mater. Interfaces},
  year    = {2026},
  volume  = {18},
  number  = {10},
  pages   = {15654--15664},
  publisher = {American Chemical Society}
}

@article{Kriegner_2016,
  author  = {Kriegner, D. and V\'yborn\'y, K. and Olejn\'ik, K. and Reichlov\'a, H. and Nov\'ak, V. and Marti, X. and Gazquez, J. and Saidl, V. and N\v{e}mec, P. and Volobuev, V. V. and Springholz, G. and Hol\'y, V. and Jungwirth, T.},
  title   = {Multiple-stable anisotropic magnetoresistance memory in antiferromagnetic MnTe},
  journal = {Nat. Commun.},
  year    = {2016},
  volume  = {7},
  number  = {1},
  pages   = {11623},
 
}

@article{Lee_2025,
  author  = {Lee, Ji-Eun and Zhong, Yong and Li, Qile and Edmonds, Mark T. and Shen, Zhi-Xun and Hwang, Choongyu and Mo, Sung-Kwan},
  title   = {Dichotomous Temperature Response in the Electronic Structure of Epitaxially Grown Altermagnet MnTe},
  journal = {Nano Lett.},
  year    = {2025},
  volume  = {25},
  number  = {22},
  pages   = {8969--8975},
  publisher = {American Chemical Society}
}

@misc{shao_2026,
      title={Epitaxial Growth and Anomalous Hall Effect in High-Quality Altermagnetic $\alpha$-MnTe Thin Films}, 
      author={Tian-Hao Shao and Xingze Dai and Wenyu Hu and Ming-Yuan Zhu and Yuanqiang He and Lin-He Yang and Jingjing Liu and Meng Yang and Xiang-Rui Liu and Jing-Jing Shi and Tian-Yi Xiao and Yu-Jie Hao and Xiao-Ming Ma and Yue Dai and Meng Zeng and Qinwu Gao and Gan Wang and Junxue Li and Chao Wang and Chang Liu},
     journal = {arXiv},
    year={2026}
}

@article{Liu_2026,
  title = {Strain-Tunable Anomalous Hall Effect in Hexagonal MnTe},
  author = {Liu, Zhaoyu and Xu, Sijie and DeStefano, Jonathan M. and Rosenberg, Elliott and Zhang, Tingjun and Li, Jinyulin and Stone, Matthew B. and Ye, Feng and Tian, Wei and Edwards, Sarah and Cong, Rong and Pan, Siyu and Chu, Ching-Wu and Deng, Liangzi and Morosan, Emilia and Fernandes, Rafael M. and Chu, Jiun-Haw and Dai, Pengcheng},
  journal = {Phys. Rev. X},
  volume = {16},
  issue = {3},
  pages = {031033},
  numpages = {26},
  year = {2026},
  month = {Aug},
  publisher = {American Physical Society}
}

@article{Wu_2020,
    author = {Wu, Mingxing and Isshiki, Hironari and Chen, Taishi and Higo, Tomoya and Nakatsuji, Satoru and Otani, YoshiChika},
    title = {Magneto-optical Kerr effect in a non-collinear antiferromagnet Mn3Ge},
    journal = {Appl. Phys. Lett.},
    volume = {116},
    number = {13},
    pages = {132408},
    year = {2020},
    month = {04}
}

@article{Marti_2014,
   title={Room-temperature antiferromagnetic memory resistor},
   volume={13},
   ISSN={1476-4660},
 
   number={4},
   journal={Nat. Mater.},
   publisher={Springer Science and Business Media LLC},
   author={Marti, X. and Fina, I. and Frontera, C. and Liu, Jian and Wadley, P. and He, Q. and Paull, R. J. and Clarkson, J. D. and Kudrnovský, J. and Turek, I. and Kuneš, J. and Yi, D. and Chu, J-H. and Nelson, C. T. and You, L. and Arenholz, E. and Salahuddin, S. and Fontcuberta, J. and Jungwirth, T. and Ramesh, R.},
   year={2014},
   month=Jan, pages={367–374} }

@article{PhysRevX_Smejkal,
  title = {Emerging Research Landscape of Altermagnetism},
  author = {Smejkal, Libor and Sinova, Jairo and Jungwirth, Tomas},
  journal = {Phys. Rev. X},
  volume = {12},
  issue = {4},
  pages = {040501},
  numpages = {27},
  year = {2022},
  month = {Dec},
  publisher = {American Physical Society}
}

@article{Fender_2025,
  author  = {Fender, Shannon S. and Gonzalez, Oscar and Bediako, D. Kwabena},
  title   = {Altermagnetism: A Chemical Perspective},
  journal = {J. Am. Chem. Soc.},
  year    = {2025},
  volume  = {147},
  number  = {3},
  pages   = {2257--2274},
  publisher = {American Chemical Society}
}

@article{Guerrero_2026,
  author  = {Guerrero-Sanchez, J. and Ponce-Perez, R. and Hoat, D. M. and Gonzalez-Hernandez, R.},
  title   = {Emergent altermagnetism and topological response in Janus MnPSX monolayers},
  journal = {Sci. Rep.},
  year    = {2026},
  volume  = {16},
  number  = {1},
  pages   = {13056},
  publisher = {Nature Publishing Group}
}

@article{Kriegner_2017,
  title = {Magnetic anisotropy in antiferromagnetic hexagonal MnTe},
  author = {Kriegner, D. and Reichlova, H. and Grenzer, J. and Schmidt, W. and Ressouche, E. and Godinho, J. and Wagner, T. and Martin, S. Y. and Shick, A. B. and Volobuev, V. V. and Springholz, G. and Hol\'y, V. and Wunderlich, J. and Jungwirth, T. and V\'yborn\'y, K.},
  journal = {Phys. Rev. B},
  volume = {96},
  issue = {21},
  pages = {214418},
  numpages = {8},
  year = {2017},
  month = {Dec},
  publisher = {American Physical Society},

}

@article{Li_2022,
  author  = {Li, Shuaixing and Wu, Jianghua and Liang, Binxi and Liu, Luhao and Zhang, Wei and Wazir, Nasrullah and Zhou, Jian and Liu, Yuwei and Nie, Yuefeng and Hao, Yufeng and Wang, Peng and Wang, Lin and Shi, Yi and Li, Songlin},
  title   = {Antiferromagnetic $\alpha$-MnTe: Molten-Salt-Assisted Chemical Vapor Deposition Growth and Magneto-Transport Properties},
  journal = {Chem. Mater.},
  year    = {2022},
  volume  = {34},
  number  = {2},
  pages   = {873--880},
  publisher = {American Chemical Society}
}

@article{Takahashi_2025,
  author  = {Takahashi, Koichiro and Huang, Hong-Fei and Yu, Jie-Xiang and Zang, Jiadong},
  title   = {Symmetry and Minimal Hamiltonian of Nonsymmorphic Collinear Antiferromagnet MnTe},
  journal = {npj Quantum Mater.},
  year    = {2025},
  volume  = {10},
  number  = {1},
  pages   = {70},
  publisher = {Nature Publishing Group}
}

@article{Devaraj_2026,
  title = {Unlocking doping effects on altermagnetism in MnTe: Emergence of quasi-altermagnetism},
  author = {Devaraj, Nayana and Bose, Anumita and Das, Arindom and Reja, Md Afsar and Mandal, Arijit and Narayan, Awadhesh and Nanda, B. R. K.},
  journal = {Phys. Rev. B},
  volume = {113},
  issue = {10},
  pages = {104438},
  numpages = {18},
  year = {2026},
  month = {Mar},
  publisher = {American Physical Society}
}

@article{Li_2026,
  title = {First-principles study of the magnon spectrum of altermagnetic MnTe},
  author = {Li, Zhengtian and Liu, Xiaoqiang and Qiao, Zhenhua},
  journal = {Phys. Rev. B},
  volume = {113},
  issue = {14},
  pages = {144301},
  numpages = {6},
  year = {2026},
  month = {Apr},
  publisher = {American Physical Society}
}

@article{Dzian_2025,
  title = {Antiferromagnetic resonance in $\ensuremath{\alpha}\text{\ensuremath{-}}\mathrm{MnTe}$},
  author = {Dzian, J. and Kuba\ifmmode \check{s}\else \v{s}{}\ifmmode \check{c}\else \v{c}{}\'{\i}k, P. and T\'azlar\ifmmode \mathring{u}\else \r{u}{}, S. and Bia\l{}ek, M. and \ifmmode \check{S}\else \v{S}{}indler, M. and Le Mardel\'e, F. and Kadlec, C. and Kadlec, F. and Gryglas-Borysiewicz, M. and Kluczyk, K. P. and Mycielski, A. and Skupi\ifmmode \acute{n}\else \'{n}{}ski, P. and Hejtm\'anek, J. and Tesa\ifmmode \check{r}\else \v{r}{}, R. and \ifmmode \check{Z}\else \v{Z}{}elezn\'y, J. and Barra, A.-L. and Faugeras, C. and Voln\'y, J. and Uhl\'{\i}\ifmmode \check{r}\else \v{r}{}ov\'a, K. and N\'advorn\'{\i}k, L. and Veis, M. and V\'yborn\'y, K. and Orlita, M.},
  journal = {Phys. Rev. B},
  volume = {112},
  issue = {2},
  pages = {024433},
  numpages = {12},
  year = {2025},
  month = {Jul},
  publisher = {American Physical Society}
}

@article{Zhang_2020,
author = {Zhang, Jiyue and Lian, Qin and Pan, Zhiqiang and Bai, Wei and Yang, Jing and Zhang, Yuanyuan and Tang, Xiaodong and Chu, Junhao},
title = {Spin-phonon coupling and two-magnons scattering behaviors in hexagonal NiAs-type antiferromagnetic MnTe epitaxial films},
journal = {J. Raman Spectrosc.},
volume = {51},
number = {8},
pages = {1383-1389},
year = {2020}
}

@article{wu_2025,
title = {Bulk single crystal growth and magneto-transport properties of {$\alpha$}-MnTe},
journal = {Journal of Magnetism and Magnetic Materials},
volume = {627},
pages = {173106},
year = {2025},
issn = {0304-8853},
author = {Si Wu and Yanping Huang and Hao Song and Baomin Wang}
}

@article{Magnin_2012,
  title = {Monte Carlo study of magnetic resistivity in semiconducting MnTe},
  author = {Magnin, Y. and Diep, H. T.},
  journal = {Phys. Rev. B},
  volume = {85},
  issue = {18},
  pages = {184413},
  numpages = {5},
  year = {2012},
  month = {May},
  publisher = {American Physical Society}
}

@article{Khatua_2025,
   title={Magnon topology driven by altermagnetism},
   volume={112},
   ISSN={2469-9969},
   number={21},
   journal={Phys. Rev. B},
   publisher={American Physical Society (APS)},
   author={Khatua, Subhankar and Kravchuk, Volodymyr P. and Yershov, Kostiantyn V. and van den Brink, Jeroen},
   year={2025},
   month=Dec }

@article{Wang_2026,
  author    = {Yong-Kun Wang and Si Li and Shengyuan A. Yang},
  title     = {Two-Dimensional Altermagnetic Iron Oxyhalides: Real Chern Topology and Valley--Spin--Lattice Coupling},
  journal   = {Nano Lett.},
  year      = {2026},
  volume    = {26},
  number    = {2},
  pages     = {831--838},
  publisher = {American Chemical Society},
  issn      = {1530-6984}
}

@article{Nagaosa_2010,
  title = {Anomalous Hall effect},
  author = {Nagaosa, Naoto and Sinova, Jairo and Onoda, Shigeki and MacDonald, A. H. and Ong, N. P.},
  journal = {Rev. Mod. Phys.},
  volume = {82},
  issue = {2},
  pages = {1539--1592},
  numpages = {0},
  year = {2010},
  month = {May},
  publisher = {American Physical Society}
}

@article{Siddiquee2023,
  author = {Siddiquee, Hasan and Broyles, Christopher and Kotta, Erica and Liu, Shouzheng and Peng, Shiyu and Kong, Tai and Kang, Byungkyun and Zhu, Qiang and Lee, Yongbin and Ke, Liqin and Weng, Hongming and Denlinger, Jonathan D. and Wray, L. Andrew and Ran, Sheng},
  
  title = {Breakdown of the scaling relation of anomalous Hall effect in Kondo lattice ferromagnet USbTe},
  
  journal = {Nat. Commun.},
  
  year = {2023},
  
  volume = {14},
  
  number = {1},
  
  pages = {527}
}

@article{PhysRevLett.49.405,
  title = {Quantized Hall Conductance in a Two-Dimensional Periodic Potential},
  author = {Thouless, D. J. and Kohmoto, M. and Nightingale, M. P. and den Nijs, M.},
  journal = {Phys. Rev. Lett.},
  volume = {49},
  issue = {6},
  pages = {405--408},
  numpages = {0},
  year = {1982},
  month = {Aug},
  publisher = {American Physical Society}
}

@article{doi:10.1143/JPSJ.12.570,
author = {Kubo ,Ryogo},
title = {Statistical-Mechanical Theory of Irreversible Processes. I. General Theory and Simple Applications to Magnetic and Conduction Problems},
journal = {J. Phys. Soc. Jpn.},
volume = {12},
number = {6},
pages = {570-586},
year = {1957}

}

@article{Blochl1994-wp,
  title = {Projector augmented-wave method},
  author = {Bl\"ochl, P. E.},
  journal = {Phys. Rev. B},
  volume = {50},
  issue = {24},
  pages = {17953--17979},
  numpages = {0},
  year = {1994},
  month = {Dec},
  publisher = {American Physical Society}
}

@article{PhysRevB.59.1758,
  title = {From ultrasoft pseudopotentials to the projector augmented-wave method},
  author = {Kresse, G. and Joubert, D.},
  journal = {Phys. Rev. B},
  volume = {59},
  issue = {3},
  pages = {1758--1775},
  numpages = {0},
  year = {1999},
  month = {Jan},
  publisher = {American Physical Society}
}

@article{Dai_2026,
author = {Dai, Yu and Zhang, Chenyu and Zhang, Yongsen and Zhang, Zeyu and Chen, Hongliang and Liu, Liang and Ma, Tianping and Hu, Yong and Yang, Mengmeng and Wang, Shouguo},
title = {Observation of Unidirectional Anisotropic Magnetoresistance in Altermagnetic MnTe (0001) Films},
journal = {Advanced Functional Materials},
volume = {36},
number = {35},
pages = {e26026},
year = {2026}
}

@article{ernzerhof1998generalized,
  title = {Generalized Gradient Approximation Made Simple},
  author = {Perdew, John P. and Burke, Kieron and Ernzerhof, Matthias},
  journal = {Phys. Rev. Lett.},
  volume = {77},
  issue = {18},
  pages = {3865--3868},
  numpages = {0},
  year = {1996},
  month = {Oct},
  publisher = {American Physical Society},
  doi = {10.1103/PhysRevLett.77.3865},
  url = {https://link.aps.org/doi/10.1103/PhysRevLett.77.3865}
}

@article{Nirmal_2024,
author = {Rooj, Suman and Chakraborty, Jayita and Ganguli, Nirmal},
title = {Hexagonal MnTe with Antiferromagnetic Spin Splitting and Hidden Rashba–Dresselhaus Interaction for Antiferromagnetic Spintronics},
journal = {Advanced Physics Research},
volume = {3},
number = {1},
pages = {2300050},
year = {2024}
}

@article{PhysRevLett.130.036702,
  title = {Spontaneous Anomalous Hall Effect Arising from an Unconventional Compensated Magnetic Phase in a Semiconductor},
  author = {Gonzalez Betancourt, R. D. and Zub\'a\ifmmode \check{c}\else \v{c}{}, J. and Gonzalez-Hernandez, R. and Geishendorf, K. and \ifmmode \check{S}\else \v{S}{}ob\'a\ifmmode \check{n}\else \v{n}{}, Z. and Springholz, G. and Olejn\'{\i}k, K. and \ifmmode \check{S}\else \v{S}{}mejkal, L. and Sinova, J. and Jungwirth, T. and Goennenwein, S. T. B. and Thomas, A. and Reichlov\'a, H. and \ifmmode \check{Z}\else \v{Z}{}elezn\'y, J. and Kriegner, D.},
  journal = {Phys. Rev. Lett.},
  volume = {130},
  issue = {3},
  pages = {036702},
  numpages = {7},
  year = {2023},
  month = {Jan},
  publisher = {American Physical Society},
  doi = {10.1103/PhysRevLett.130.036702},
  url = {https://link.aps.org/doi/10.1103/PhysRevLett.130.036702}
}

@article{dudarev,
  title = {Electron-energy-loss spectra and the structural stability of nickel oxide:  An {LSDA+U} study},
  author = {Dudarev, S. L. and Botton, G. A. and Savrasov, S. Y. and Humphreys, C. J. and Sutton, A. P.},
  journal = {Phys. Rev. B},
  volume = {57},
  issue = {3},
  pages = {1505--1509},
  numpages = {0},
  year = {1998},
  month = {Jan},
  publisher = {American Physical Society},
  doi = {10.1103/PhysRevB.57.1505},
  url = {https://link.aps.org/doi/10.1103/PhysRevB.57.1505}
}

@article{Das_2026,
  title = {Effect of symmetry breaking on altermagnetism in CrSb and formation of fragmented nodal curves},
  author = {Das, Arindom and Mandal, Arijit and Devaraj, Nayana and Nanda, B. R. K.},
  journal = {Phys. Rev. B},
  volume = {113},
  issue = {18},
  pages = {184443},
  numpages = {11},
  year = {2026},
  month = {May},
  publisher = {American Physical Society},
  doi = {10.1103/w47d-pjxf},
  url = {https://link.aps.org/doi/10.1103/w47d-pjxf}
}

@article{PhysRevB.111.184407,
  title = {Spontaneous anomalous Hall effect in two-dimensional altermagnets},
  author = {Sheoran, Sajjan and Dev, Pratibha},
  journal = {Phys. Rev. B},
  volume = {111},
  issue = {18},
  pages = {184407},
  numpages = {9},
  year = {2025},
  month = {May},
  publisher = {American Physical Society},
  doi = {10.1103/PhysRevB.111.184407},
  url = {https://link.aps.org/doi/10.1103/PhysRevB.111.184407}
}

@article{MOSTOFI2008685,
title = {wannier90: A tool for obtaining maximally-localised Wannier functions},
journal = {Computer Physics Communications},
volume = {178},
number = {9},
pages = {685-699},
year = {2008},
issn = {0010-4655},
doi = {https://doi.org/10.1016/j.cpc.2007.11.016},
url = {https://www.sciencedirect.com/science/article/pii/S0010465507004936},
author = {Arash A. Mostofi and Jonathan R. Yates and Young-Su Lee and Ivo Souza and David Vanderbilt and Nicola Marzari}
}

@article{zhao_2026_theory,
  author = {Yufei Zhao and Saswata Mandal and Chao-Xing Liu and Binghai Yan},
  title = {Emergent Anomalous Hall Effect from Surface States in the Altermagnet {MnTe} Thin Films},
  journal = {arXiv preprint},
  year = {2026},
  eprint = {2603.12259}
}

@article{Junior_2023,
  title = {Sensitivity of the MnTe valence band to the orientation of magnetic moments},
  author = {Faria Junior, Paulo E. and de Mare, Koen A. and Zollner, Klaus and Ahn, Kyo-hoon and Erlingsson, Sigurdur I. and van Schilfgaarde, Mark and V\'yborn\'y, Karel},
  journal = {Phys. Rev. B},
  volume = {107},
  issue = {10},
  pages = {L100417},
  numpages = {5},
  year = {2023},
  publisher = {American Physical Society}
}

@article{wang_2026_InP_MnTe,
  title = {Phase and electronic structure control of altermagnetic $\ensuremath{\alpha}\text{\ensuremath{-}}\mathrm{MnTe}$ thin films via epitaxy on InP (111)},
  author = {Wang, Yuzhe and Su, Shiwu and Ge, Min and Ran, Pengxu and Li, Tongrui and Li, Xianglin and Liao, Sen and Liang, Yu and Wen, Jiaqin and Yao, Jianghao and Sun, Jiexiong and Yan, Haoyuan and Xu, Rui and Cui, Shengtao and Sun, Zhe and Feng, Donglai and Jiang, Juan},
  journal = {Phys. Rev. B},
  volume = {113},
  issue = {20},
  pages = {205139},
  numpages = {7},
  year = {2026},
  month = {May},
  publisher = {American Physical Society},
  doi = {10.1103/vwgc-ltkz},
  url = {https://link.aps.org/doi/10.1103/vwgc-ltkz}
}

@article{Mandal2025,
  title = {Deterministic role of chemical bonding in the formation of altermagnetism: Reflection from the correlated electron system NiS},
  volume = {112},
  ISSN = {2469-9969},
  DOI = {10.1103/sm63-1dcx},
  number = {1},
  journal = {Physical Review B},
  publisher = {American Physical Society (APS)},
  author = {Mandal,  Arijit and Das,  Arindom and Nanda,  B. R. K.},
  year = {2025},
  month = {July}
}

@article{Orlova2024,
  author  = {Orlova, N. N. and Avakyants, A. A. and Timonina, A. V. and Kolesnikov, N. N. and Deviatov, E. V.},
  title   = {Crossover from Relativistic to Non-Relativistic Net Magnetization for MnTe Altermagnet Candidate},
  journal = {JETP Lett.},
  volume  = {120},
  number  = {5},
  pages   = {360--366},
  year    = {2024},
  issn    = {1090-6487}
}

@article{Efrem_2005,
  title   = {Low-temperature neutron diffraction study of MnTe},
  journal = {J. Magn. Magn. Mater.},
  volume  = {285},
  number  = {1},
  pages   = {267--271},
  year    = {2005},
  issn    = {0304-8853},
   author  = {J. B. C. {Efrem D'Sa} and P. A. Bhobe and K. R. Priolkar and A. Das and S. K. Paranjpe and R. B. Prabhu and P. R. Sarode}
}

@article{Zhou_2018,
    author = {Zhou, Chunsheng and Wang, Jianfang and Liu, Xin and Chen, Fengying and Di, Youying and Gao, Shengli and Shi, Quan},
    title = {Magnetic and thermodynamic properties of $\alpha$, $\beta$, $\gamma$ and $\delta$-MnO2},
    journal = {New J. Chem.},
    volume = {42},
    number = {11},
    pages = {8400-8407},
    year = {2018},
    month = {06},
    issn = {1144-0546}
}

@article{Belashchenko_2025,
  title = {Giant Strain-Induced Spin Splitting Effect in MnTe, a g-Wave Altermagnetic Semiconductor},
  author = {Belashchenko, K. D.},
  journal = {Phys. Rev. Lett.},
  volume = {134},
  issue = {8},
  pages = {086701},
  numpages = {6},
  year = {2025},
  publisher = {American Physical Society} 
}

@inproceedings{Wasscher_1969,
  title={Electrical transport phenomena in MnTe, an antiferromagnetic semiconductor},
  author={J. D. Wasscher},
  year={1969}
}

@misc{bey_2026,
      title={Conductivity scaling of the anomalous Hall effect in the altermagnetic semiconductor alpha-MnTe}, 
      author={Sara Bey and Shelby S. Fields and Nicholas G. Combs and Bence G. Márkus and Jiashu Wang and Liam Schmidt and Lincoln Curtis and Allecia Dodd-Noble and Alexander Poulin and Syed Mohammad Shahed and Resham Regmi and Mariia Holub and Phillipe Ohresser and Arun Bansil and Swastik Kar and Haile Ambaye and Valeria Lauter and László Forró and Cory D. Cress and Joseph C. Prestigiacomo and Nirmal Ghimire and Alberto de la Torre and Steven P. Bennett and Xinyu Liu and Badih A. Assaf},
      year={2026},
      journal = {arXiv preprint},
      eprint={2603.00242},
      archivePrefix={arXiv},
      primaryClass={cond-mat.str-el},
     }

\newpage

\end{document}

\def\scititle{Supplementary Information for the manuscript -- Berry Curvature Driven Transport in Silicon-Compatible Altermagnetic $\alpha$-MnTe Thin Films}

\title{\bfseries \boldmath \scititle}
\author[1]{\fnm{Rajib} \sur{Sarkar}}
\author[1]{\fnm{Subhransu Kumar} \sur{Negi}}
\author[2,3]{\fnm{Arindom} \sur{Das}}
\author[2,3]{\fnm{Arijit} \sur{Mandal}}
\author[1]{\fnm{Pankaj} \sur{Bhardwaj}}
\author[1]{\fnm{Sohini} \sur{Guin}}
\author[1]{\fnm{Aryaman} \sur{Das}}
\author[1]{\fnm{Naresh} \sur{Shyaga}}
\author[1]{\fnm{Kartick} \sur{Biswas}}
\author[1]{\fnm{Laxmipriya} \sur{Nanda}}
\author[1]{\fnm{Pavan} \sur{Nukala}}
\author*[2,3]{\fnm{B. R. K. } \sur{Nanda}}\email{nandab@iitm.ac.in}
\author*[1]{\fnm{Dhavala} \sur{Suri}}\email{dsuri@iisc.ac.in}

\affil[1]{\orgdiv{Centre for Nanoscience and Engineering}, \orgname{Indian Institute of Science}, \orgaddress{\street{Bangalore}, \city{Bengaluru}, \postcode{560012}, \state{Karnataka}, \country{India}}}

\affil[2]{\orgdiv{Condensed Matter Theory and Computational Lab}, \orgname{Department of Physics}, \orgaddress{\street{IIT Madras }, \city{Chennai}, \postcode{600036}, \state{Tamil Nadu}, \country{India}}}

\affil[3]{\orgdiv{Center for Atomistic Modelling and Materials Design}, \orgaddress{\street{IIT Madras }, \city{Chennai}, \postcode{600036}, \state{Tamil Nadu}, \country{India}}}

\maketitle

\newpage
\begin{appendices}

\section{In-situ structural evolution (RHEED)}\label{secA1}

During the growth of MnTe thin films, in situ reflection high energy electron diffraction (RHEED) images were recorded at different time intervals within the MBE chamber to monitor structural evolution. The initial images correspond to the Si(111) substrate, which exhibits a predominantly streaky pattern with some superimposed spots. This combination indicates a clean, well-ordered crystalline surface, confirming that the pre-growth preparation is effective and that no oxidized or amorphous regions are present.

\begin{figure}[h!]
\centering
\includegraphics[width=0.8\linewidth]{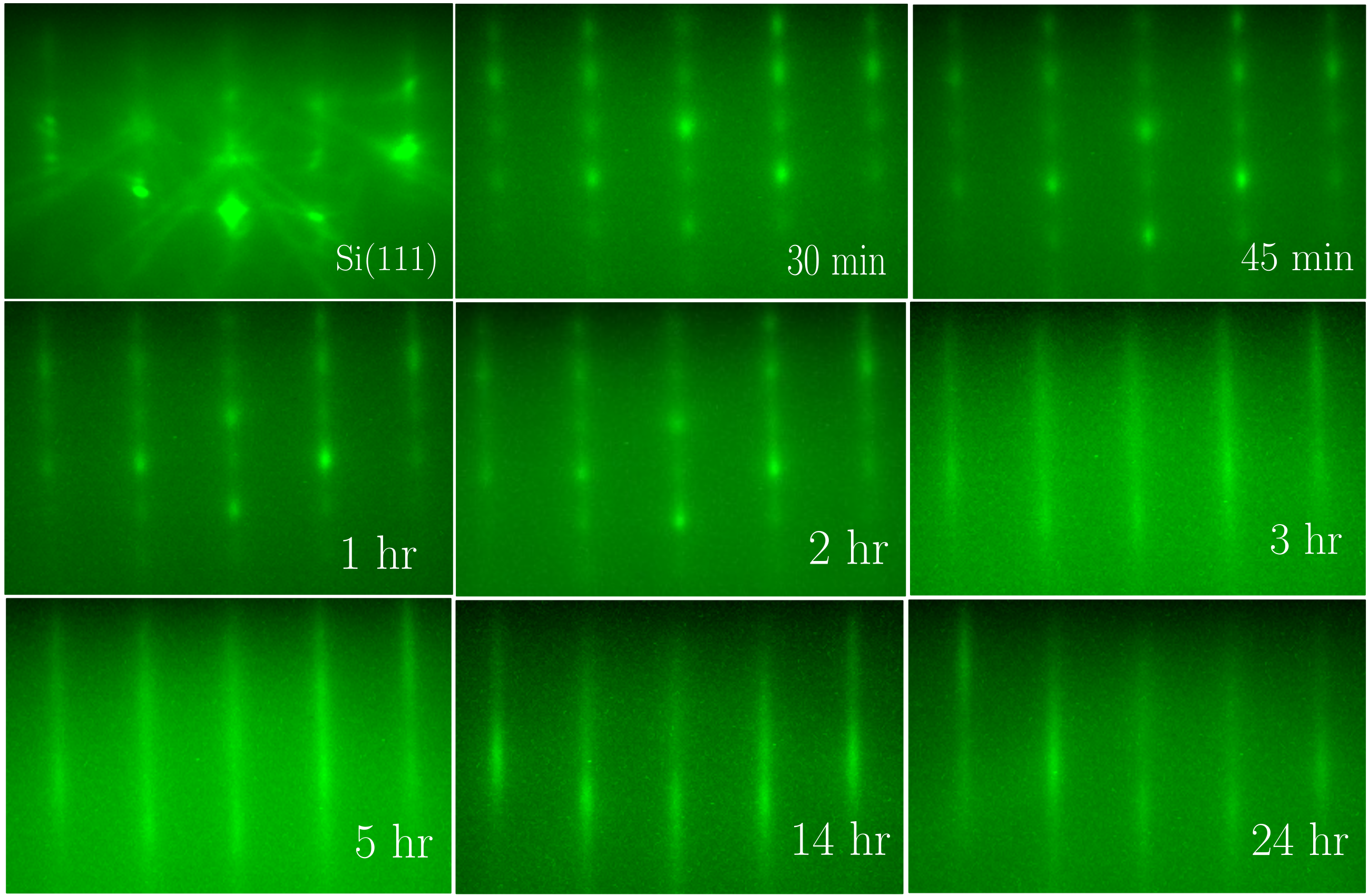}
\caption{Evolution of RHEED patterns during film growth at different time intervals}
\label{fig:Rheed_evolution}
\end{figure}

As the growth progresses, the RHEED patterns evolve. In the early stages, the images remain largely streaky but develop additional spotty features, indicative of a transition toward three-dimensional (3D) growth. This behavior is attributed to interfacial strain arising from the significant lattice mismatch between MnTe and Si(111). To mitigate these effects, the growth rate was intentionally kept very low, allowing sufficient time for adatom diffusion and surface reorganization.

With continued deposition, the RHEED patterns gradually recover a more pronounced streaky character, reflecting an improvement in surface smoothness and structural ordering. After approximately 3 hours of growth, well-defined streaky patterns are observed. For thicker films, deposition was extended up to 24 hours. Despite the presence of strain, the RHEED pattern quality does not degrade with increasing thickness. This suggests that while strain influences the interface during the initial stages, it does not adversely affect the subsequent growth process, ultimately enabling the formation of high-quality MnTe thin films.

\newpage
\section{Structural characterization (XRD).}
We performed $2\theta$ XRD measurements on three samples of varying thickness. The diffraction patterns show that the (002) and (004) reflections from MnTe are prominent and consistent across all samples, indicating that the films are preferentially oriented along the (00c) direction. However, in contrast to the XRD pattern in the main figure, MT33 shows an additional (110) peak with lower intensity than the other prominent peaks. This sample has an FWHM of 0.34$^{\circ}$, larger than that of the MT60 samples, which have an FWHM of 0.2$^{\circ}$. In the main Fig. 1(d-f), we discuss the highly epitaxial film using $\phi$-scanning and RSM. This suggests that both samples are highly qualitative, but MT60 is more so, with no additional peaks other than the (00c) peaks.

\begin{figure}
    \centering
    \includegraphics[width=0.6\linewidth]{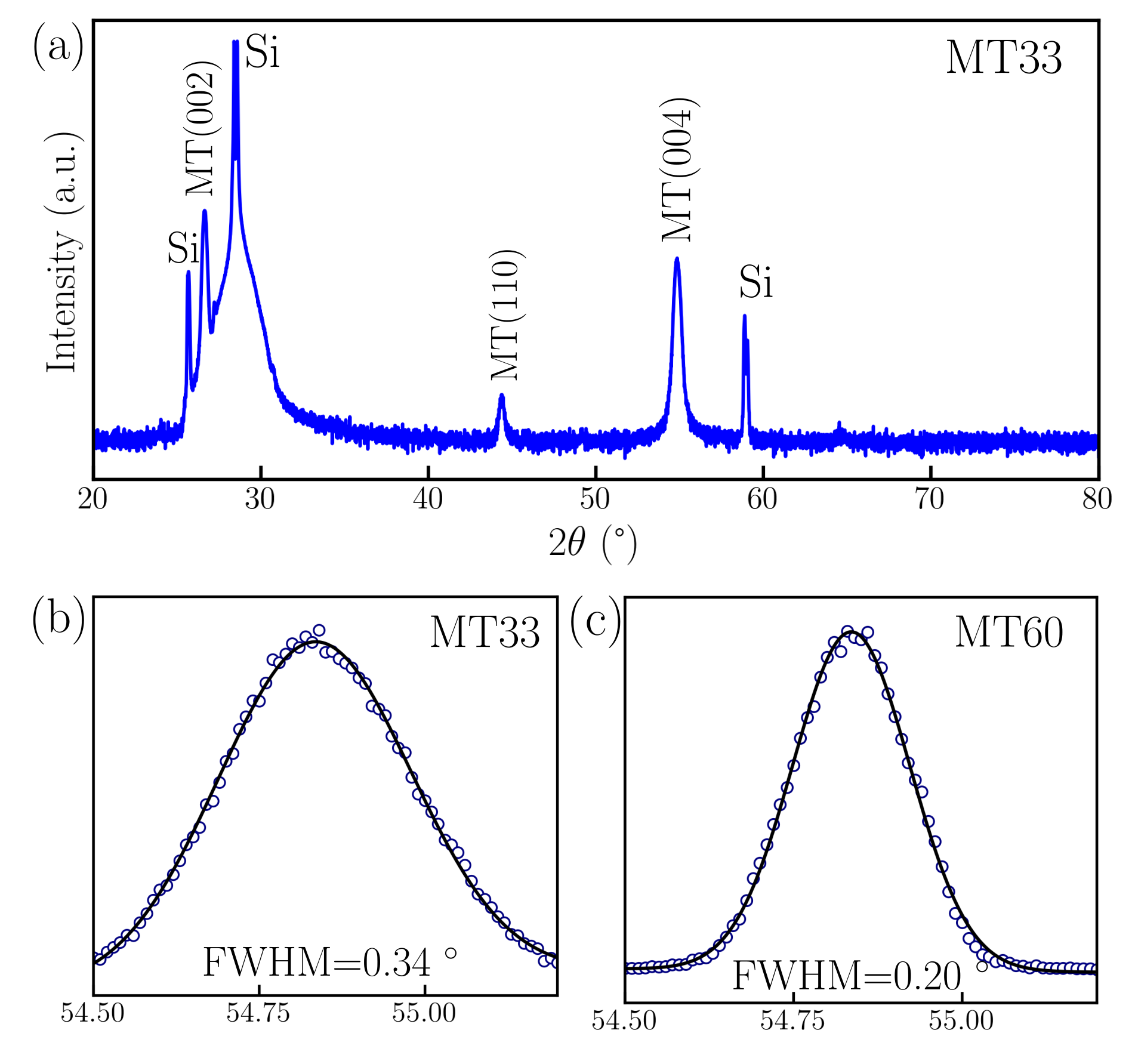}
    \caption{(a) Full XRD spectrum of MT34. (b, c) XRD spectra of the (004) MnTe peak, showing FWHM values of 0.34$^\circ$ and 0.20$^\circ$ for the MT33 and MT60 samples, respectively.
 }
    \label{fig:XRD_MT33_MT60}
\end{figure}
 The extracted lattice parameter from the RHEED and XRD has been shown below. 
\begin{table}[htbp]
\color{black}
    \centering
    \caption{Lattice parameters and $c/a$ ratio of MnTe grown on different substrates.}
    \label{tab:lattice_parameters}
    \begin{tabular}{lccc}
    \hline
    Substrate & $a$ (\AA) & $c$ (\AA) & $c/a$ \\
    \hline
    MnTe/InP(111) & 4.178 & 6.68 & 1.59 \\
    MnTe/GaAs(111) & 4.17 & 6.69 & 1.6 \\
    bulk MnTe & 4.16 & 6.71 & 1.61 \\
    \hline
    \hline
    MnTe/Si(111) [MT47] & 4.08 & 6.688 & 1.6390 \\
    MnTe/Si(111) [MT33] & 4.082 & 6.692 & 1.6390 \\
    MnTe/Si(111) [MT34] & 4.084 & 6.688 & 1.6376 \\
    \hline
    \end{tabular}
\end{table}
However, the c value of our sample is close to the bulk value but differs from the in-place lattice parameter, giving rise to intrinsic strain in the sample. Recent study suggests that MnTe is very sensitive to its lattice parameter, resulting in variation in sample properties such as band gap, band splitting\cite{Belashchenko_2025}, different Neel vector orientation, or presence or absence of AHE. Hence, we used a$=4.08\AA $ and c$=6.69\AA$ (experimental values) for calculations to avoid uncertainty.

\newpage

\section{Atomic Force Microscopy }
\begin{figure}[h!]
    \centering
    \includegraphics[width=0.5\linewidth]{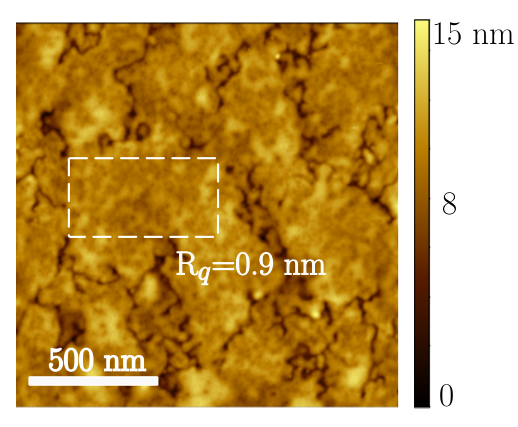}
    \caption{Atomic force microscopic image of MT47 sample}
    \label{fig:SI_AFM}
\end{figure}

To examine the surface morphology, we performed AFM imaging (Fig. \ref{fig:SI_AFM}). As the sample thickness increases, the size of the granules increases consistently. The roughness ($R_q$) over an area 500~$nm^2$ remains $\leq$~1nm for all our samples. High lattice mismatch during growth causes increasing strain over time, resulting in rougher surfaces over large areas. 

\section{Transmission Electron Microsocopy Imaging Analysis}

\begin{figure}[h!]
\centering
\includegraphics[width=1.0\linewidth]{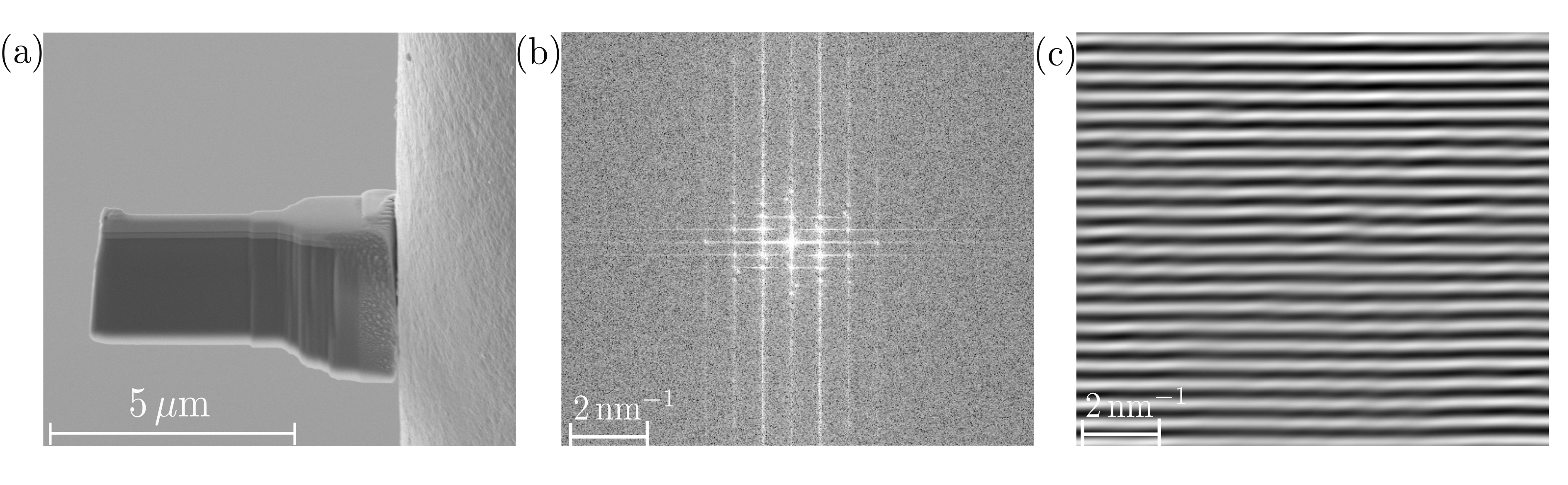}
\caption{ TEM analysis (a) Overview of the MnTe (MT60) thin film and the Si(111). SEM image of prepared lamella of the MnTe thin film at (b) Selective area Diffraction pattern (SAED) of MnTe layer,(c)Inverse FFT image obtained from the corresponding HRTEM micrograph.}
\label{fig:SI_TEM}

\end{figure}

This section shows the TEM image analysis of MnTe thin film. Fig.~\ref{fig:SI_TEM} represents overview of the MnTe thin film on Si(111). The SAED pattern of MnTe indicates the Epitaxial growth of the same as that of the RHEED pattern.

We measured the d-spacing in different regions closer and away  from the surface. We find that the d-spacing is the same indicating that there is no MnTe$_2$ or any other secondary phase formation. The contrast in the image acquistion, shows the same.

\begin{figure}[h!]
\centering
\includegraphics[width=0.8\linewidth]{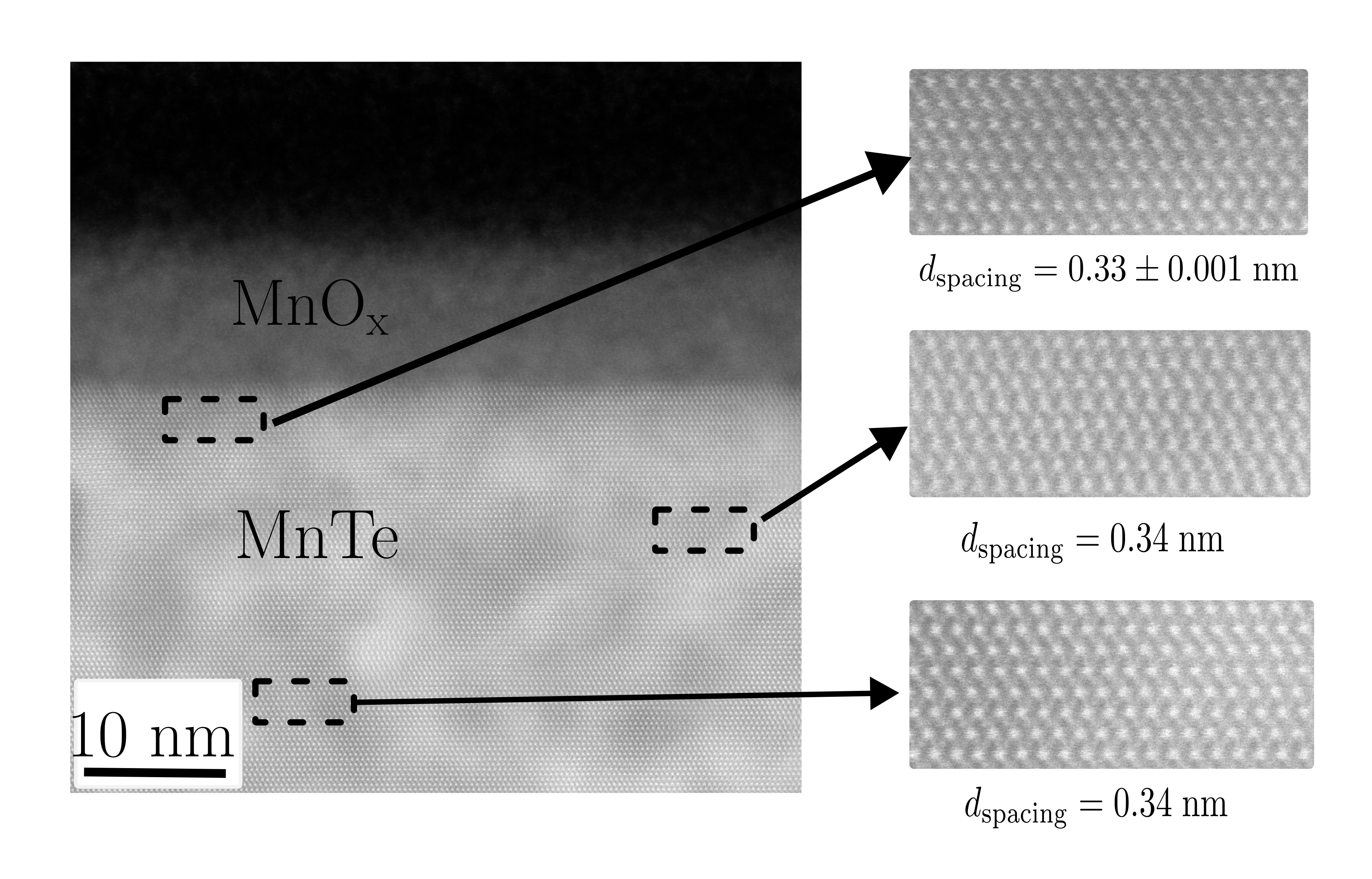}
\caption{d-spacing in different regions of MT60 thin film.}
\label{fig:TEM_regions}
\end{figure}

\newpage
\section{X-ray Photoemission Spectroscopy (XPS) of the samples}
\begin{figure}[h!]
\centering
\includegraphics[width=1\linewidth]{SI_XPS.pdf}
\caption{Deconvolved High-resolution core XPS spectrum of (a) Mn-2p and (b) Te-3d }
\label{fig:SI_XPS}
\end{figure}
To probe the chemical environment, XPS measurements were performed on both Mn and Te core levels. The Mn 2$p_{3/2}$ spectrum can be well described by a combination of Mn$^{2+}$ and Mn$^{3+}$ components, indicating a mixed-valence state. Such behavior may arise from local deviations in stoichiometry, defect states, or interfacial charge transfer effects, and can have important implications for both magnetic ordering and transport properties. The Te 3$d$ spectrum  shows a dominant contribution from Te bound to Mn, along with a weaker component associated with Te--O bonding, indicative of minor surface oxidation. Importantly, the relative intensity of the oxide component remains small, suggesting that the bulk of the film retains the expected chemical integrity. Taken together, these structural and spectroscopic results confirm the formation of high-quality MnTe thin films with well-defined crystallographic orientation and predominantly stoichiometric bonding, while also highlighting subtle electronic and magnetic inhomogeneities that may play a role in the observed transport phenomena.

\section{Transport Measurements for additional samples }

\begin{figure}[h!]
    \centering
    \includegraphics[width=0.8\linewidth]{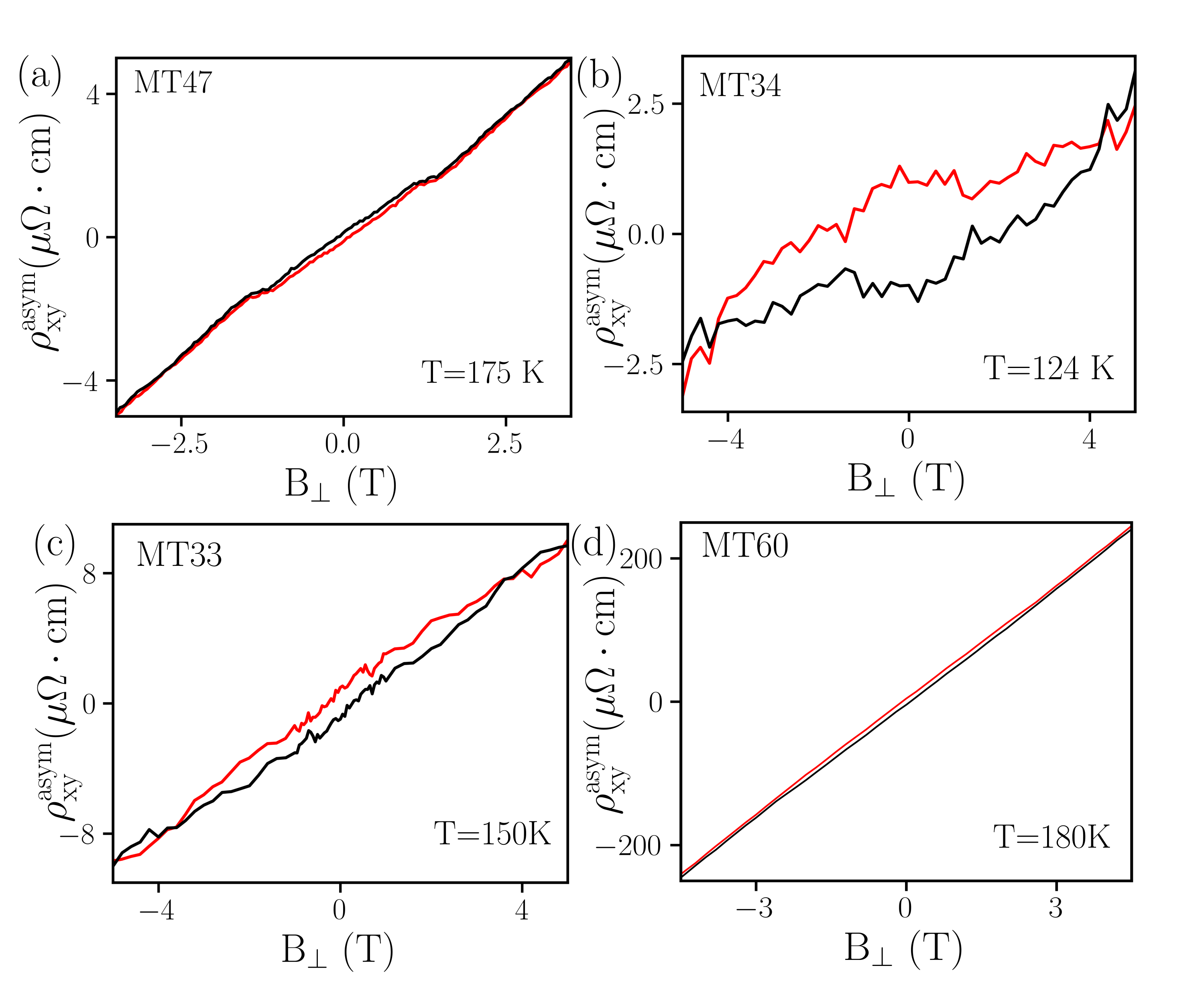}
    \caption{Hall Resistance vs Magnetic field in different set of samples:(a)MT47(10nmn), (b)MT34(20nm), (c)MT33(50nm) and (d)MT60(200nm)}
    \label{fig:Hall_SI}
\end{figure} 
We perform magnetic-field-dependent resistance measurements on different samples, where MT33, MT34, and MT47 carry an (110) peak, but MT60 does not, similar to MT59 but with a different thickness. All samples show clear hysteresis, suggesting that a small deviation in sample quality does not eliminate AHE. Additionally, the sample shows hysteresis in both thick and thin samples, suggesting that AHE contains both surface and bulk contributions. In addition, we show the thickness dependence of Hall resistivity in Fig.~\ref{fig:scaling}~(a), which indicates that it does not monotonically increase with thickness but saturates beyond a certain thickness. This clearly indicates that AHE has both surface and bulk dependence.

\subsection{In-Plane Magnetotransport Measurements}
\begin{figure}[h!]
    \centering
    \includegraphics[width=0.6\linewidth]{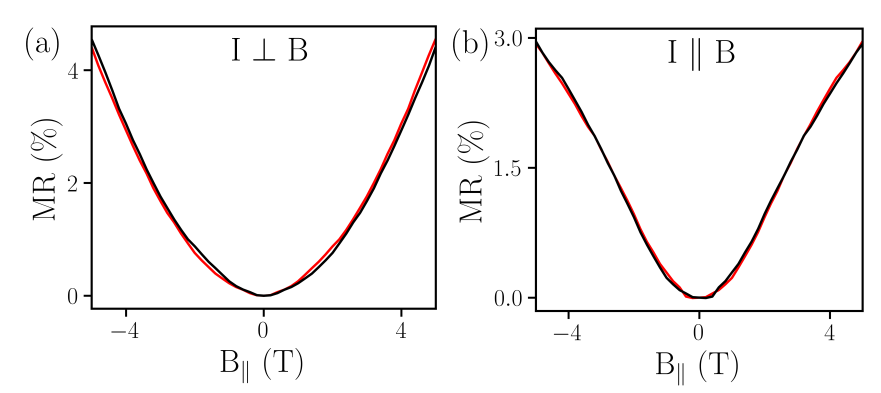}
    \caption{In-plane magnetotransport measurements of the MT34 sample for (a) I $\perp$ B and
(b) I $\parallel$ B at 175 K}
    \label{fig:MT34_IP_175K}
\end{figure}
Here, we present the magnetoresistance data in the presence of an in-plane magnetic field. However, we did not observe any hysteresis curve in the magnetoresistance, unlike the out-of-plane magnetic field, which shows that the anomalous Hall effect (AHE) is present only in the out-of-plane configuration (as shown in main Fig. 4). Although the magnetic field is aligned in-plane, there is no hysteresis in the in-plane magnetoresistance, confirming that there is no prominent ferromagnetic contribution in the ab-plane. This suggests the antiferromagnetic (AFM) nature of MnTe and anisotropy in the system.

\section{Anomalous Hall Effect}
\begin{figure}[h!]
    \centering
    \includegraphics[width=0.8\linewidth]{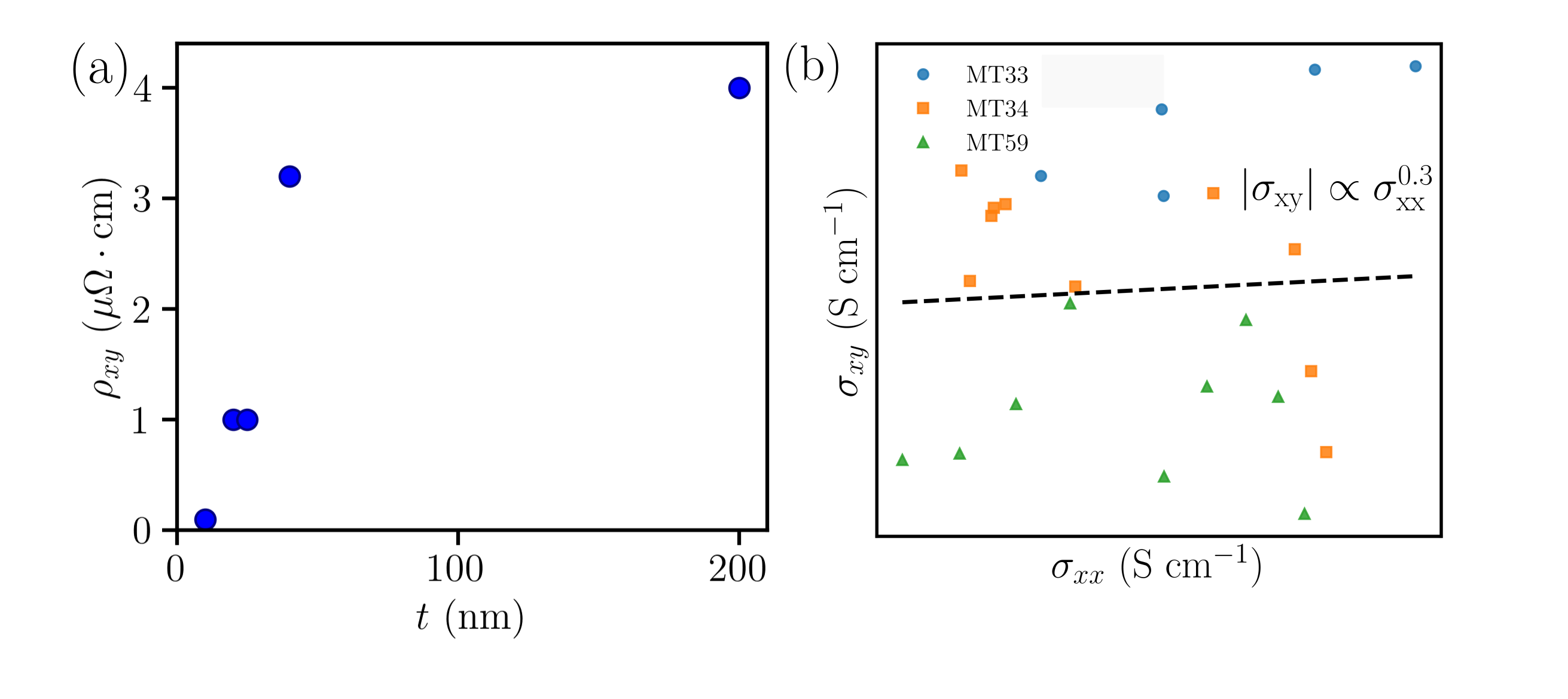}
    \caption{(a) Resistivity vs thickness, (b) Scaling relation  between $\sigma_{xy}$ vs $\sigma_{xx}$.}
    \label{fig:scaling}
\end{figure}

In Fig.~\ref{fig:scaling}~(a), we have plotted the Hall resistivity $\rho_{xy}$  as a function of the thickness of the films. Initially, the hall resistivity increases with sample thickness. This got saturated after a certain thickness, which suggests that the primary contribution comes from the surface and secondary contribution comes from  the surface and secondary contribution comes from bulk inner  layers close to the surface.

Anomalous Hall conductivity (AHC) versus longitudinal conductivity  is  supported by prior experimental studies demonstrating that the absence of scaling between $\sigma_{xy}$ and $\sigma_{xx}$ is indicative of a dominant intrinsic contribution($\sigma_{xy} $ is independent of $\sigma_{xy} $). Our results show no clear systematic scaling between \(\sigma_{xy}\) and \(\sigma_{xx}\), which is consistent with a significant intrinsic contribution to the observed AHE. Although this analysis does not by itself uniquely establish a Berry-curvature origin, when considered together with the magnetization measurements and first-principles calculations, it supports a predominantly intrinsic Berry-curvature contribution.

The theoretical calculations are carried out on a slab of few (4-5) unit cell thickness; the slab is exposed to vacuum.  The calculated AHC lies in the range 50  $\mu S$ -- 120  $\mu S$. This unit differs from the bulk unit in Scm$^{-1}$; since the reciprocal space in defined in two dimensions. These values are of the same order as those reported earlier \cite{zhao_2026_theory}.    Experimental values are obtained on samples of 10-200 nm thickness, with AHC values of $\approx 0.001~ \mathrm{S}$cm$^{-1}$ - $ 0.1~\mathrm{S}$ cm$^{-1}$, which translates to  10$^{-2} \mu S$ -- 10$^{-1} \mu S$, which is in agreement with reported experimental results \cite{Liu_2026}. Since the model unit cell used for DFT calculations  are very different from real samples used in experiment it may not be appropriate to quantitatively compare the transport data such as AHC. We also emphasize the fact that a minor shift in Fermi level can change the AHC by orders of magnitude Fig. 3 (c,f).

\section{Mobility}

\begin{figure}[h!]
    \centering
    \includegraphics[width=0.8\linewidth]{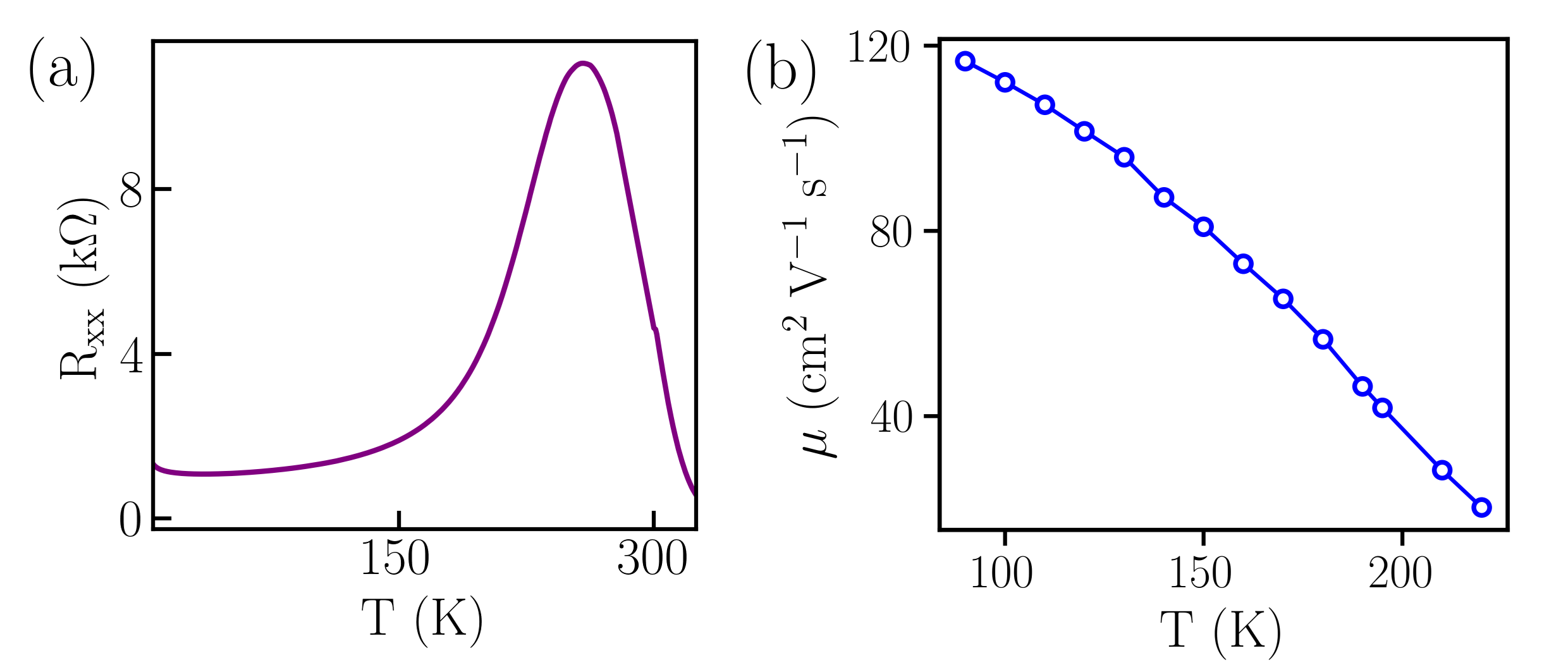}
    \caption{(a) Resistance vs temperatue and  (b) mobility vs temperature for sample MT47}
    \label{mobility_T}
\end{figure}

To determine mobility, we performed linear fitting and extracted the parameter. From the linear Hall slope of Hall measurements, we extract the carrier mobility ($\mu$)[Fig.~\ref{mobility_T}~(b)]. The mobility decreases systematically with increasing temperature, from $\sim$116 to $\sim$20 $\text{cm}^2/\text{V}\cdot\text{s}$ between 100 K and 230 K, consistent with enhanced phonon scattering at elevated temperatures. These values remain relatively high for a semiconductor film\cite{bey_2026,Wasscher_1969}. 

In table~\ref {tab:mobility}, we show the mobility of different samples. This shows that mobility is higher in the thinner sample than in the thicker sample.The relatively high mobility may also indicate a contribution from surface or interface conduction channels (as the hybridization of bulk and surface state got enhanced with thicker film,which reduces the mobility), consistent with features observed in the temperature-dependent resistance.This indicates that the surface or interface state gives rise to the higher mobility in semiconducting MnTe.

%\bibliography{references.bib}
\begin{table}[htbp]
    \centering
    \caption{Temperature-dependent mobility of the MnTe samples at 150 K.}
    \label{tab:mobility}
    \begin{tabular}{c  c}
        \hline
        Sample ID & $\mu$(cm$^2$ V$^{-1}$)  \\
        \hline
        MT47-10nm & 80.86 \\
        MT30-10nm  & 62 \\
        MT33-50nm  & 22.37 \\
        MT34-20nm & 12.1 \\
        MT59-25nm & 20.9 \\
        MT60-200nm & 12.48 \\
        \hline
    \end{tabular}
\end{table}

\section{Planar Hall Effect Measurements }
To examine the presence of magnetic anisotropy in the $xy$-plane, we performed planar Hall measurements at room temperature under an applied magnetic field of 0.45~T. The transverse resistance $R_{xy}$ exhibits a clear sinusoidal dependence on the in-plane rotation angle, indicating the presence of antiferromagnetic anisotropy within the $xy$-plane.

Notably, the observed periodicity is $2\pi$, rather than the conventional $\pi$ periodicity typically expected in planar Hall systems. This deviation suggests an unconventional origin of the planar Hall response \cite{Zhou2025,Wang2025}, which does not arise from ferromagnetic ordering or spin–orbit-coupling-driven mechanisms.
\begin{figure}[h!]
\centering
\includegraphics[width=0.6\linewidth]{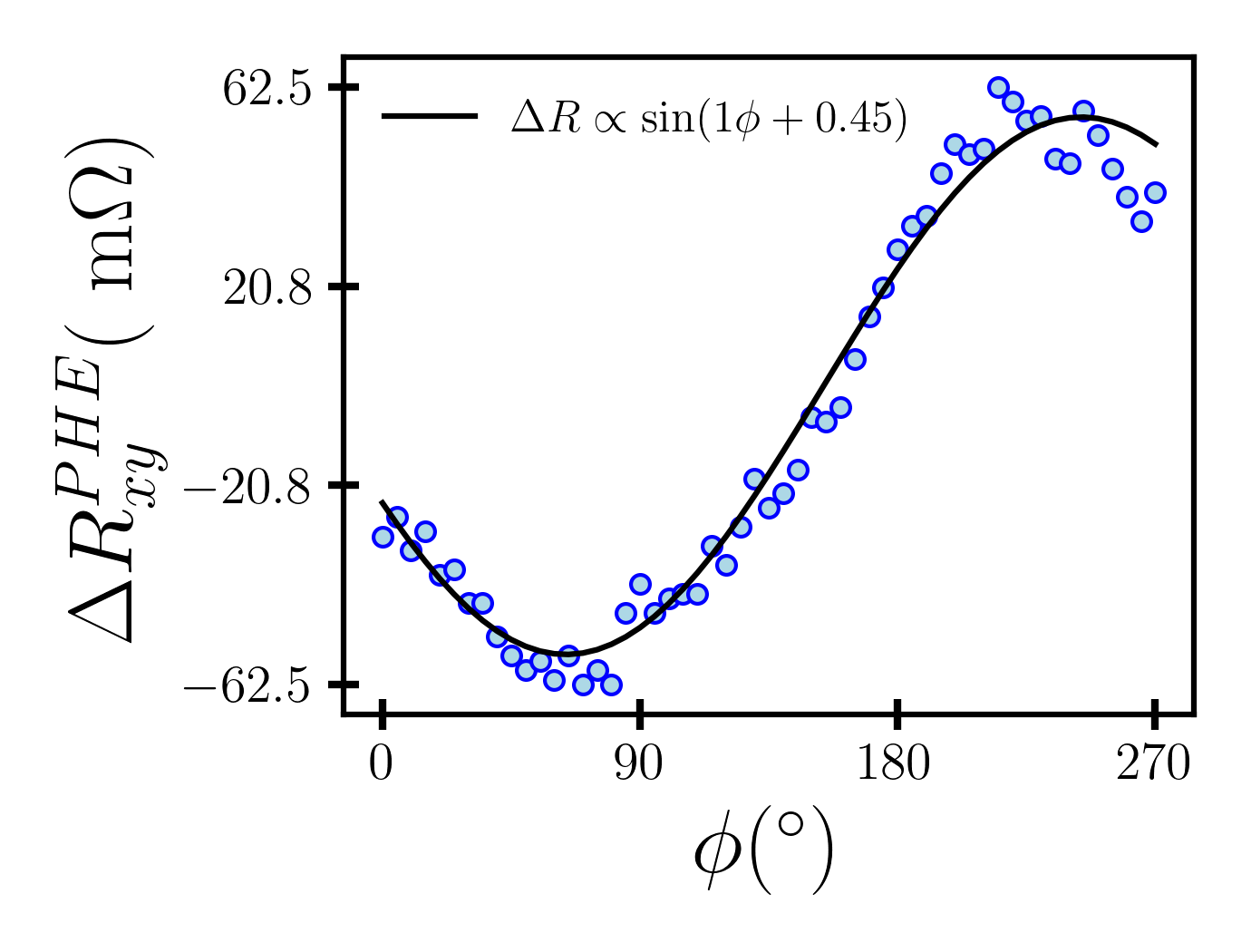}
\caption{ Existence of the planar Hall effect with $2\pi$ periodicity.}
\label{fig:PHE}
\end{figure}

\section{Effect of Sample Quality }

Interestingly, we find that improved sample quality correlates with an increase in the transition temperature. As shown in Fig.~1(g), the MT59 sample exhibits a narrower FWHM than MT33, indicating superior crystalline quality. In MT59, the transition temperature closely matches values reported in the literature for samples grown on InP substrates. These results suggest that the N\'eel temperature increases systematically with improving sample quality. Furthermore, the bulk sample exhibits a transition near 307~K, close to the N\'eel temperature observed in the strained thin film\cite{Betancourt_2023}. This indicates that while the transition temperature depends weakly on strain, sample quality exerts a comparatively stronger (higher-order) influence. However, this does not change the presence of AHE in different quantitative samples (as shown in Fig.~\ref{fig:Hall_SI})
 
\begin{figure}[h!]
    \centering
    \includegraphics[width=0.5\linewidth]{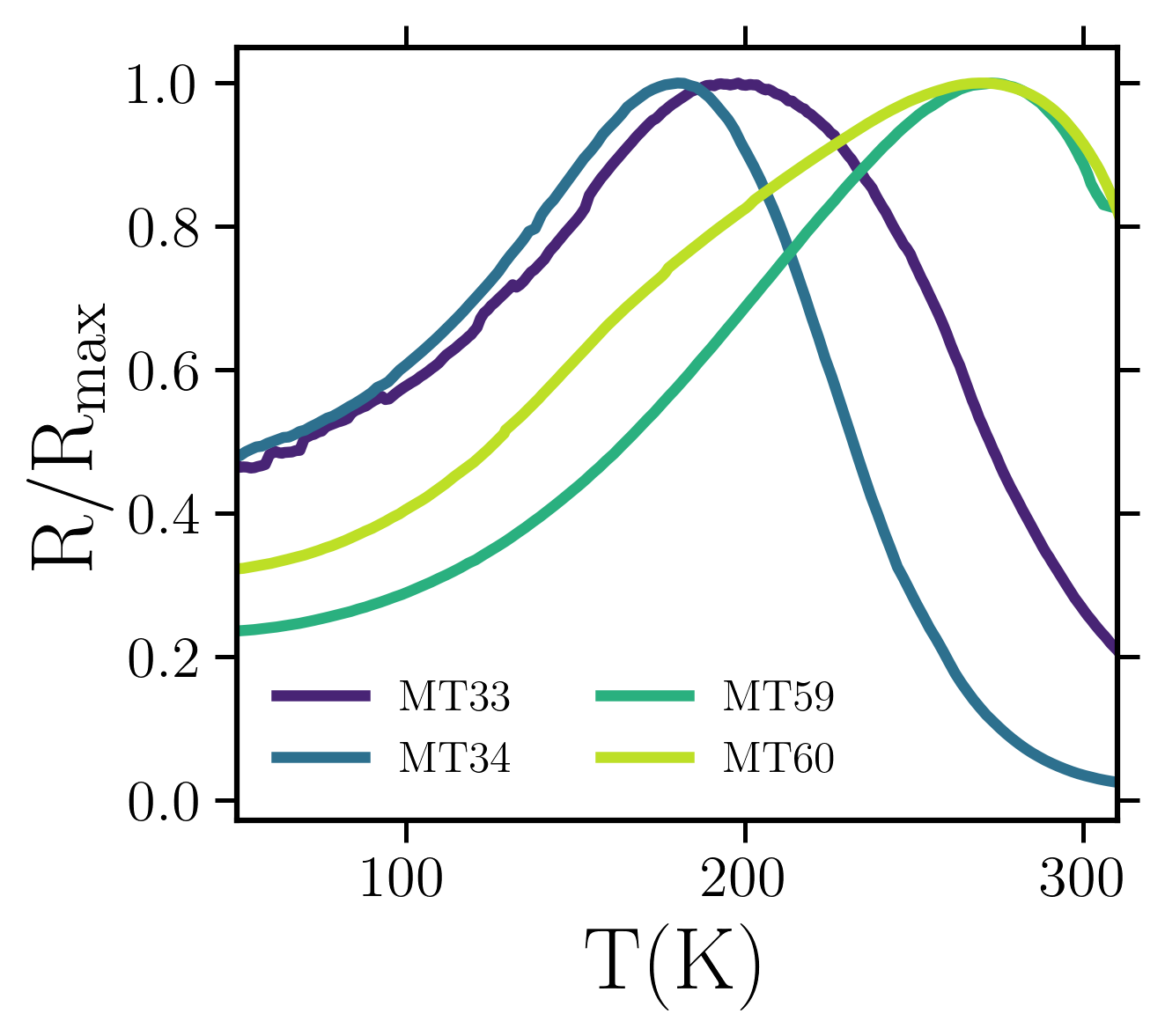}
    \caption{Resistance–temperature curves for different samples. Samples MT33 and MT34 show an additional peak corresponding to the (110) planes, whereas MT59 and MT60 do not show this peak.
}
    \label{fig:RT_diff_samples}
\end{figure}

However, the c value of our sample is close to the bulk value.  But the in-place lattice parameter differs from that of bulk implying an intrinsic strain in the sample. Recent study suggests that MnTe is very sensitive to its lattice parameter, resulting in variation in sample properties such as band gap, band splitting\cite{Belashchenko_2025}, different Neel vector orientation, or presence or absence of AHE. Hence, we used a$=4.08\AA $ and c$=6.69\AA$ (experimental values) for calculations as well.  

\begin{table}[htbp]
    \centering
    \caption{XRD peak positions observed for different MnTe samples.}
    \label{tab:xrd_peaks}
    \begin{tabular}{c c c}
        \hline
        \textbf{Sample ID} & \textbf{(00$l$) Peaks} & \textbf{(110) Peaks}  \\
        \hline
        MT33 & present  & present   \\
        MT34 & present &  present  \\
        MT47 & present &  present \\
        MT59 & present &  absent  \\
        MT60 & present &  absent  \\
        \hline
    \end{tabular}
\end{table}
\newpage

\section{Data Processing for $R_{xx}$}
Here, we have presented the data processing protocol step by step. This is intended to show that the temperature fluctuation at the sample is negligible. In any case, the contribution due to temperature fluctuation can be corrected for as well.  
\begin{figure}[h!]
\centering
\includegraphics[width=0.6 \linewidth]{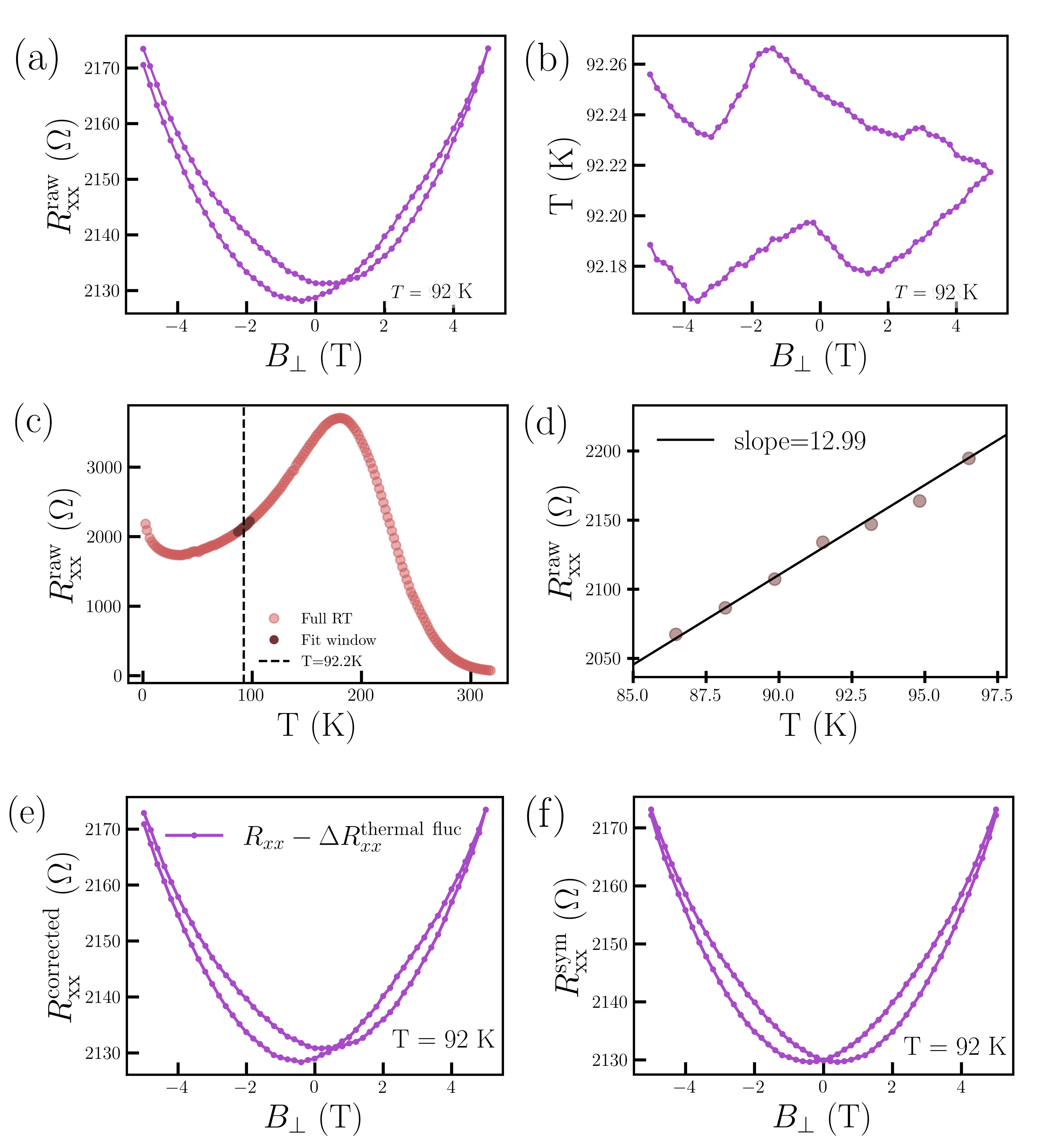}
\caption{(a) Raw data (b) Temperature fluctuation as a function of magnetic field during the process of acquiring raw data. (c) Resistance versus temperature at zero magnetic field (d) Resistance versus temperature over a temperature range 85 K to 98 K showing linear variation. (e) Raw data of resistance versus magnetic field corrected for the small temperature fluctuation $\approx$ 10 mK during the entire field sweep. (f) Symmetrized resistance with respect to magnetic field of the data corrected for temperature fluctuation.}
\label{fig:Data_Processing}
\end{figure}
Addtitionally,in the Fig~\ref{fig:T_variation}, we clearly show that the temperature variation throughout the experiment is minimal, indicating that the presence of AHE in transport data is not due to the temperature fluctuation.

\begin{figure}[h!]
    \centering
    \includegraphics[width=0.8\linewidth]{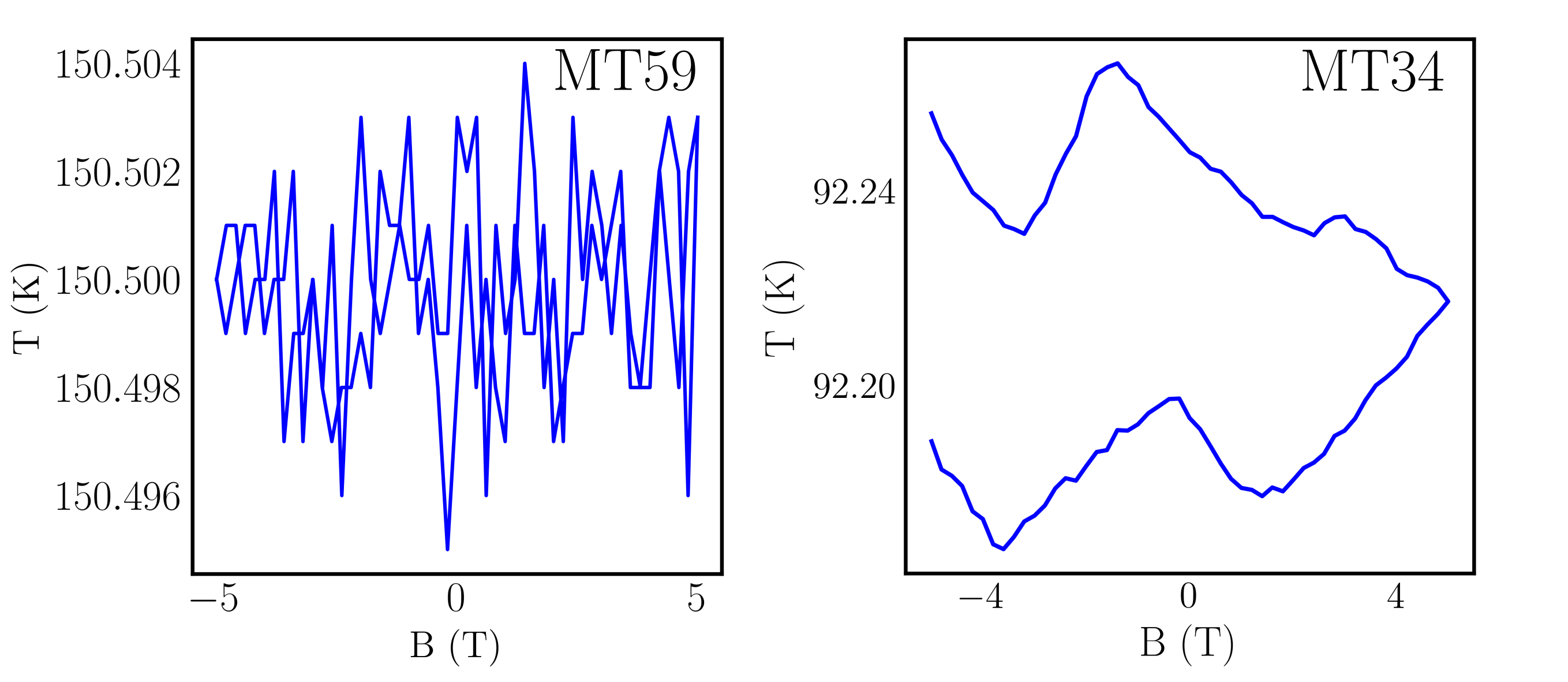}
    \caption{Temperature variations in the entire magnetic field sweeps for data acquired for the two samples (a) MT59 is the sample presented in the main manuscript. (b) MT34 is the data presented in the SI.}
    \label{fig:T_variation}
\end{figure}

\section{Magnetization measurements (VSM).}
\begin{figure}[h!]
\centering
\includegraphics[width=1.0 \linewidth]{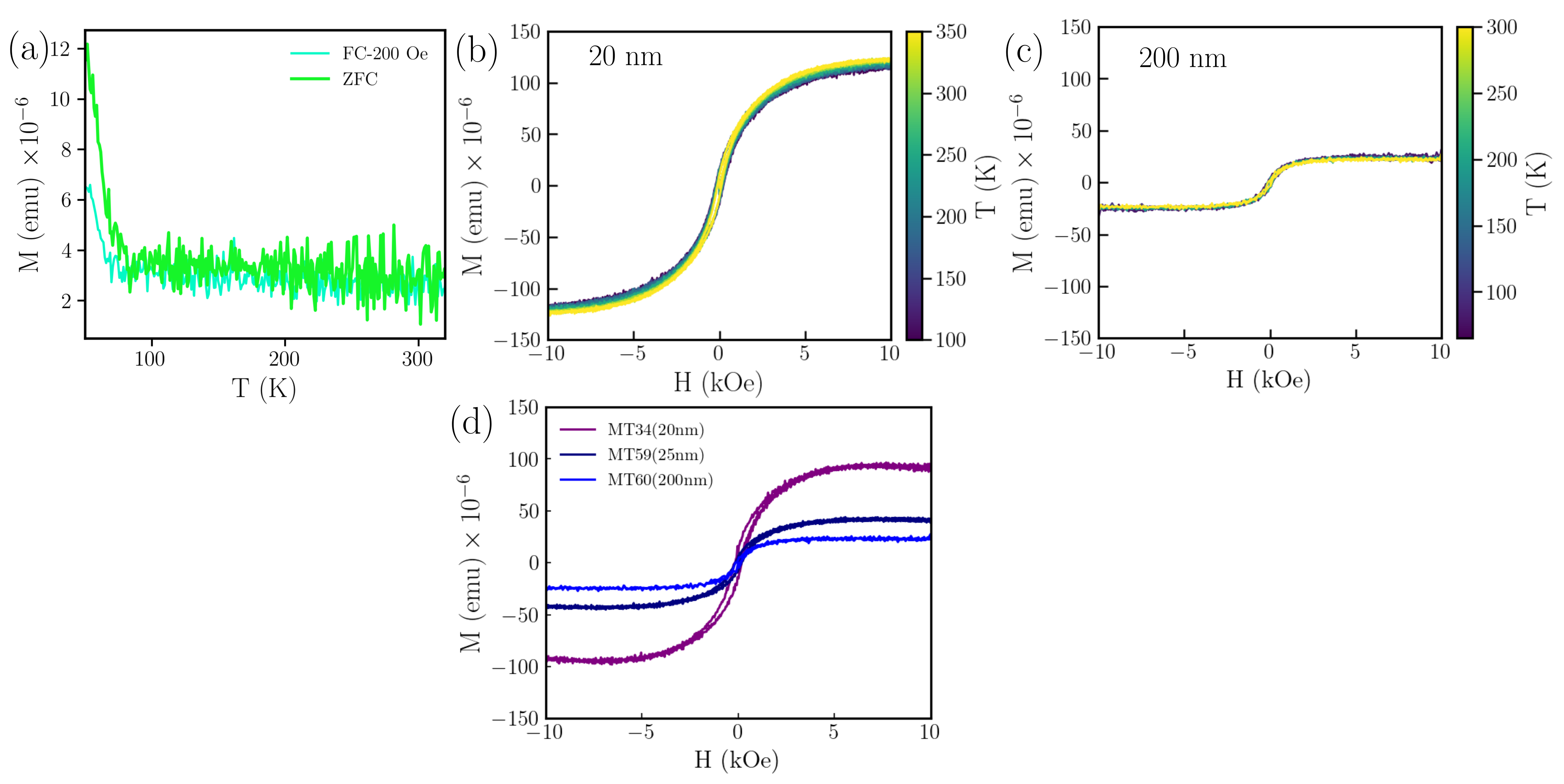}
\caption{(a) Temperature-dependent magnetization measurements, with a field (200~Oe) in the plane of the sample. (b–c) Magnetic field–dependent magnetization (M–H) curves measured at different temperatures, with the magnetic field applied in the plane of the sample, for the 20 nm and 200 nm thick samples, respectively.(d) M–H curves of different samples measured at 200 K.
}
\label{fig:MT}
\end{figure}

To confirm the absence of intrinsic magnetization in the sample and to verify that the observed anomalous Hall effect originates from Berry-curvature-driven mechanisms, temperature-dependent and thickness-dependent magnetization measurements were performed using a VersaLab VSM. 

Below approximately 80~K, the magnetization increases abruptly, resembling a ferromagnetic transition. This anomaly, however, does not necessarily signal a conventional ferromagnetic phase transition. Several mechanisms, both intrinsic and extrinsic to the MnTe films, can account for it. Additionally, we performed field-dependent magnetization measurements on samples with different thicknesses (20 and 200 nm) but identical sample areas, as shown in Fig. L15(b–c).
. The 200 nm sample shows a saturation moment approximately one order of magnitude lower than that of the 20 nm sample, despite its substantially larger film volume. This inverse scaling with thickness is inconsistent with a bulk ferromagnetic or ferrimagnetic phase, which would be expected to produce a moment that increases with film volume; instead, it supports a predominantly surface/interfacial origin for the measured magnetic signal.

However, recent literatures explain the possible reason for the origin of this anomaly in the MH curve.One experimentally shows  that in altermagnets, symmetry permits a weak spin--orbit coupling that gives rise to a finite net magnetization, reflecting a crossover between relativistic and non-relativistic contributions to the magnetic moment~\cite{Orlova2024}. A closely related precedent supports this interpretation: in a previous study reporting a similar low-temperature anomaly, low-temperature neutron diffraction ruled out both the emergence of an additional magnetic phase and any accompanying structural transition, thereby excluding a conventional change in the antiferromagnetic structure~\cite{Efrem_2005}. Formation of MnTe$_2$ ($T_N \approx 87$~K) was also excluded as a possible source of spurious signal. Instead, the anomaly was traced to a temperature-dependent variation of the $c/a$ lattice ratio below $\sim$75~K, which modifies the Mn--Mn exchange interactions by strengthening intraplanar ferromagnetic coupling while weakening interplanar antiferromagnetic coupling---a magnetoelastic effect rather than a true structural or magnetic phase transition. Because this conclusion is directly supported by neutron diffraction, it provides strong precedent for attributing the anomaly observed here to single-phase MnTe rather than to a secondary phase or genuine long-range ferromagnetic order.

Extrinsic contributions cannot be fully excluded, however. Native surface oxidation of manganese-based chalcogenides commonly produces thin, mixed-valence MnO$_x$ layers, which are known to exhibit parasitic magnetic transitions in the 80--100~K range and could superimpose an additional signal onto the intrinsic response~\cite{Zhou_2018}.

\section{Insights from electronic structure calculations}

\subsection{Electronic structure of bulk MnTe}

\begin{figure}[htbp!]
    \centering
    \includegraphics[width=0.7\linewidth]{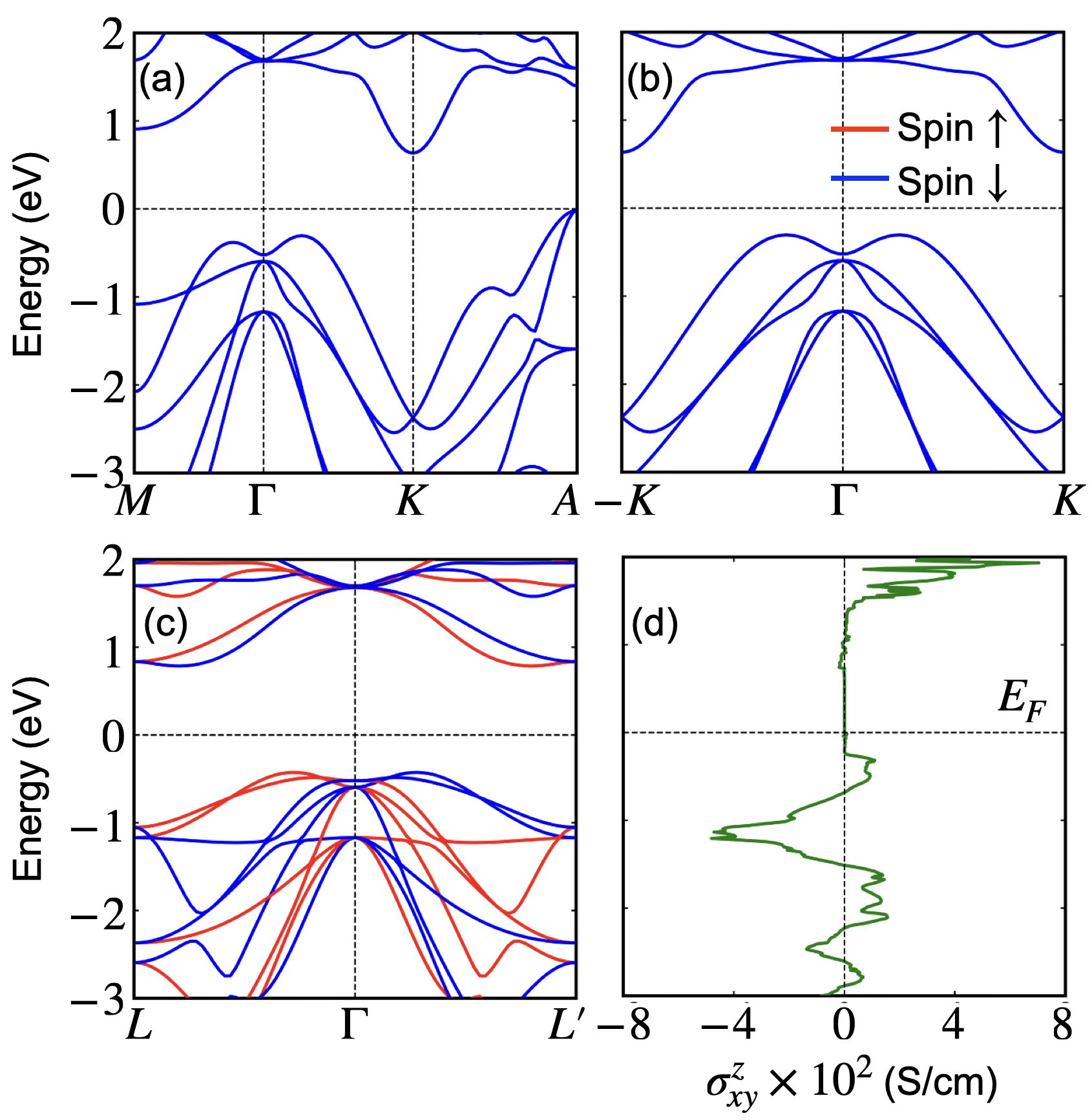}
    \caption{Non-relativistic spin polarized band structure and AHC of bulk MnTe. (a)-(c) represent the band structure along various $k$-path. In (a) and (b), the $k$-path along which the band is plotted lies in the nodal plane, resulting in a perfect antiferromagnetic band. In the figures, the spin $\uparrow$ bands are not visible due to the spin degeneracy of the band structure. (c) represents the altermagnetic band structure. (d) represents the AHC in bulk MnTe when the spin quantization axis is oriented along the $y$ direction. $E_F$ denotes the Fermi energy level. The symmetry of the compound allows only $z$ component of the AHC.}
    \label{fig:bulk_MnTe}
\end{figure}

\begin{figure}
     \centering
     \includegraphics[width=0.7\linewidth]{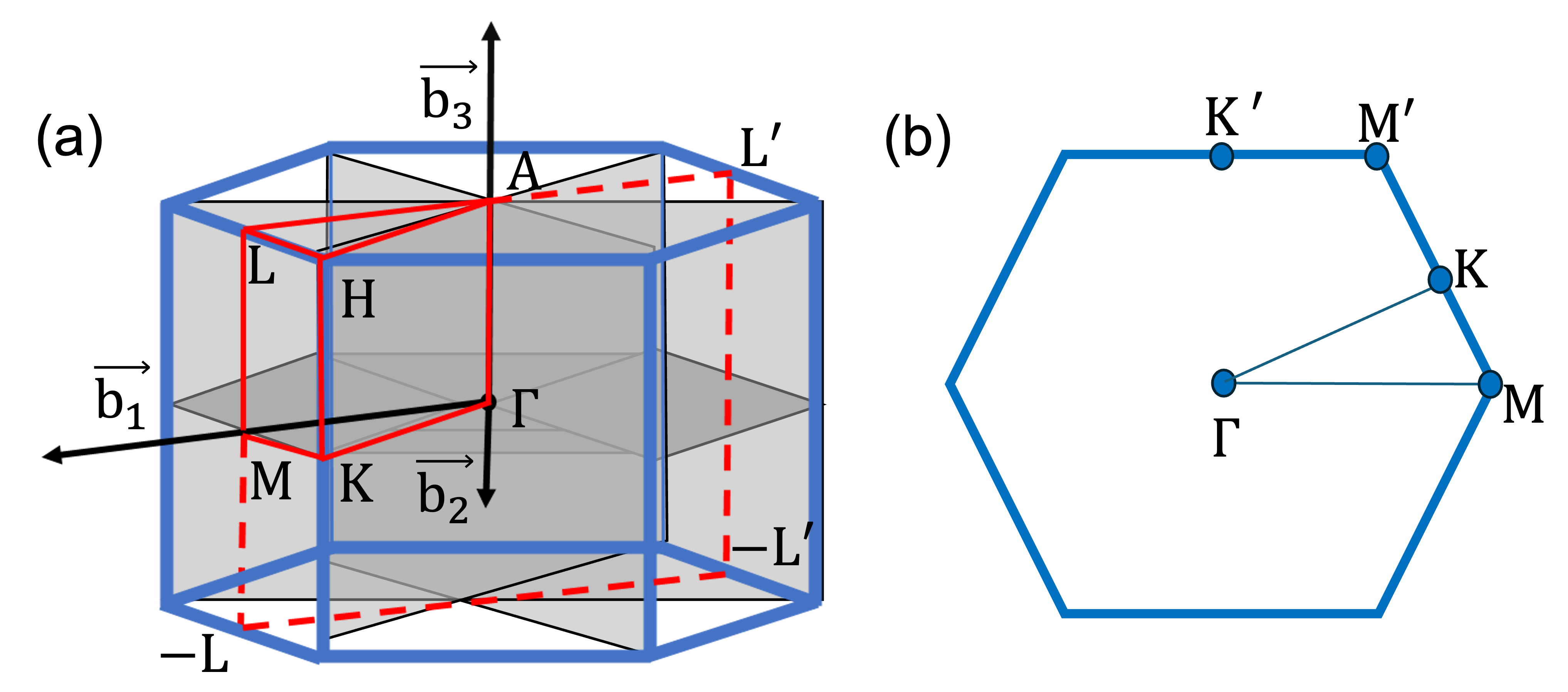}
     \caption{(a) The hexagonal BZ of bulk MnTe with the high symmetry points and four nodal planes. The red-dashed rectangle, represents one of the widely studied altermagnetic planes, where the momentum dependent spin splitting is one of the maximum. (b) The hexagonal BZ for the MnTe film grown along $z$ axis  with the high symmetry points.}
     \label{fig:BZ}
 \end{figure}

 Through Fig. \ref{fig:bulk_MnTe} we analyze the magnetic structure of the bulk MnTe. Along the k-path lying on the nodal planes \ref{fig:BZ}, the subband Krammers spin degeneracy is intact [see upper panel of Fig. \ref{fig:bulk_MnTe}], leading to zero momentum-dependent spin splitting. As we move away from the nodal planes [see \ref{fig:bulk_MnTe}-(c)], the degeneracy is broken due to momentum-dependent spin splitting, which is a characteristic feature of altermagnetism. The AHC calculated for bulk MnTe, which prefer spin quantization along $y$ axis, is shown in Fig. \ref{fig:bulk_MnTe}. Due to perfect spin compensation, the AHC is found to be zero at the Fermi level.

\subsection{Symmetry analysis of the MnTe film} 
\label{Symmetry_analysis}
The bulk MnTe belongs to the space group $P 6_3/mmc$ and exhibits A-type antiferromagnetic spin orientation. The crystallographic symmetry elements in this space group can be divided into two halving subgroups $\boldsymbol{G}-\boldsymbol{H}$ and $\boldsymbol{H}$ \cite{Mandal2025, PhysRevX_Smejkal}. The symmetry elements are as follows:

$\boldsymbol{G}-\boldsymbol{H}$ =  $\{C_{2z}t_{0, 0, 1/2}$, $C_{2(2x+y)}t_{0, 0, 1/2}$, $C_{2(x+2y)}t_{0, 0, 1/2}$, $C_{2(x-y)}t_{0, 0, 1/2}$, $M_{2z}t_{0, 0, 1/2}$, $M_{2x+y}t_{0, 0, 1/2}$, $M_{x+2y}t_{0, 0, 1/2}$, $M_{x-y}t_{0, 0, 1/2}$, $C_{6z}^{\pm}t_{0, 0, 1/2}$, $S_{6z}^{\pm}t_{0, 0, 1/2}\}$

$\boldsymbol{H}$ = $\{ I, C_{3z}^{\pm}, S_{3z}^{\pm}, P, C_{2x}, C_{2y}, C_{x+y}, M_{x}, M_{y}, M_{x+y} \}$

While $\boldsymbol{G}-\boldsymbol{H}$ contains the symmetry elements connecting the opposite spin sublattices of this altermagnetic compound, $\boldsymbol{H}$ contains the symmetry elements connecting the same spin sublattices. In the films, as the growth direction is $z$, the periodicity along the $z$ direction is killed at the surface. As a result, all symmetries that involve inversion and/or translation with respect to the $z$ axis are absent. Hence, there exists no symmetries in the subgroup $\boldsymbol{G}-\boldsymbol{H}$ to maintain the altermagnetism. However, there exist six symmetries ($I, C_{3z}^{\pm}, S_{3z}^{\pm}, M_{x}, M_{y}, M_{x+y}$) that connects the same spin sublattices. The absence of a symmetry connection between the surface Mn layer and the neighboring Mn layer, even though they belong to opposite spin sublattices, leads to an uncompensated magnetization involving these two layers. 

\subsection{Magnetic ground state and MAE calculations}

To examine the ground state of magnetic MnTe films, we carried out total energy calculations for FM and A-type AFM ordering for three configurations: config-I, where top surface is Te terminated with excess Te capping at the top [see Fig. 3-(a) of main text], config-II is where both the top and bottom surfaces are Te terminated with excess Te capping [see fig. 3-(d) of main text], and last one config-III is simply Te terminated MnTe films [see Fig. \ref{fig:AHC_no_cap_Te}-(a)].  As the Table \ref{tab:energy_comparison} suggests the A-type antiferromagnetic order is always stable for all the configurations.
\begin{table}[ht!]
\centering
\caption{Comparison of total energies per formula unit (in eV) for the FM and AFM configurations of three different configurations.}
\label{tab:energy_comparison}
    
\begin{tabular}{c|c|c}
    \hline
    Structure & FM (eV) & AFM (eV) \\
    \hline
    Config-I   & 0 & -0.240 \\
    Config-II  & 0 & -0.200 \\
    Config-III & 0 & -0.234 \\
    \hline
    \end{tabular}
\end{table}

Our magnetocrystalline anisotropy calculations demonstrated that for config-I and II, the $z$-axis becomes the easy axis, while for config-III, which is Te-terminated like bulk MnTe, with vacuum, there is spin polarization along the plane or $x$-axis. For this calculation, we have used an energy cutoff of $500$ eV for the plane-wave basis set with a $20 \times 20 \times 1$ $k$-mesh.

\begin{table}[ht!]
\centering
\caption{Comparison of total energies of the slab for spin polarizations along the $x$, $y$, $z$-axis of three different configurations.}
\label{tab:MAE_comparison}
    
\begin{tabular}{c|c|c|c}
    \hline
    Structure & x (meV) & y (meV) & z (meV)\\
    \hline
    Config-I  & 0.5440 & 0.5080 & 0 \\
    Config-II & 0.6350 & 0.6680 & 0 \\
    Config-III & 0 & 0.0120 & 3.6720 \\
    \hline
    \end{tabular}
\end{table}

\subsection{AHC in MnTe films}
\begin{figure} [ht!]
    \centering
    \includegraphics[width=0.7\linewidth]{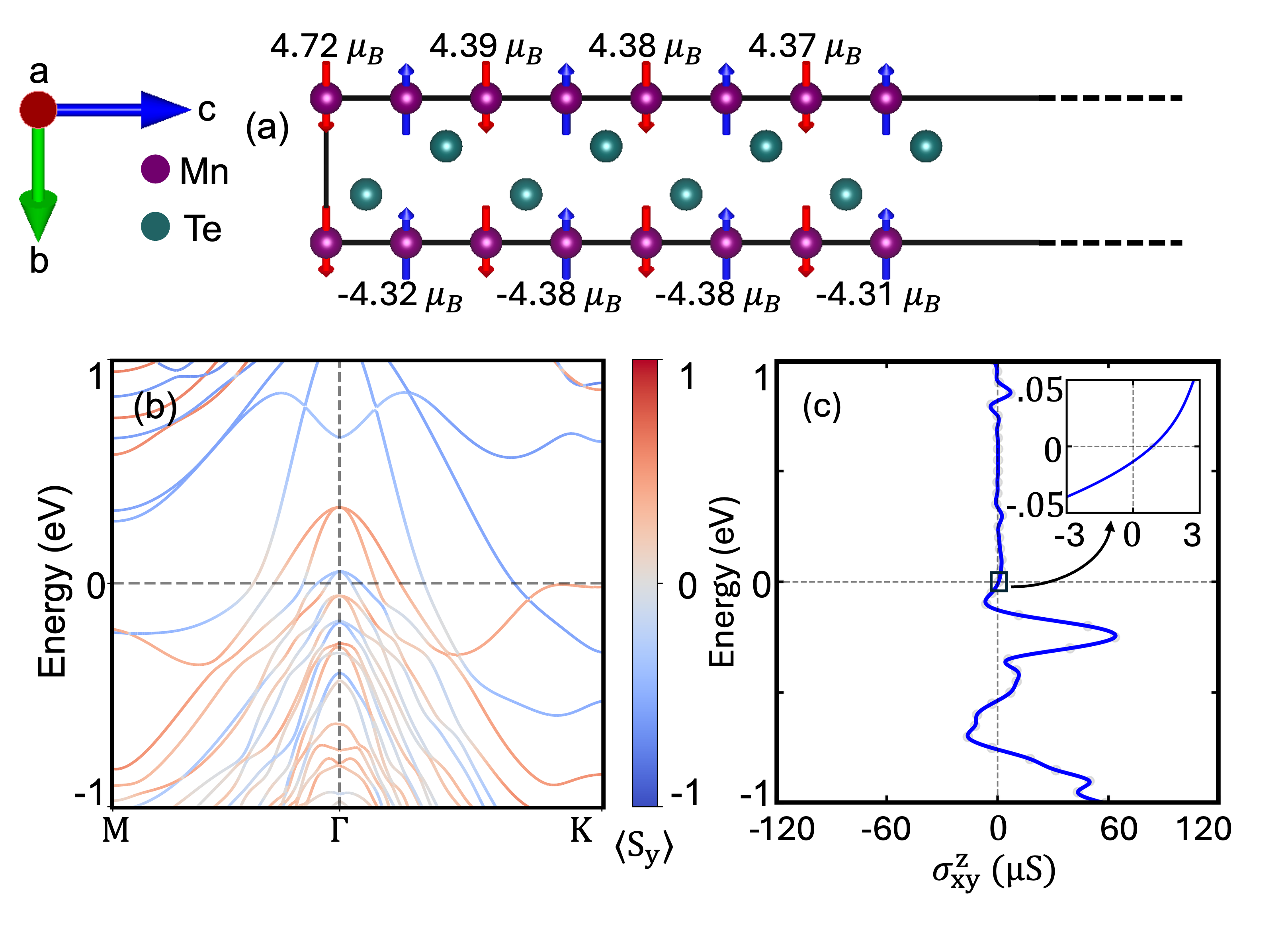}
    \caption{The schematics of the MnTe film with Te termination is shown with the individual magnetic moment (in $\mu_{B}$) of Mn atoms in (a) with spin ($\langle S_y \rangle$) projected DFT+SOC band in (b) and AHC in (c). The inset in (c) shows presence of weak but finite AHC at the Fermi level. The change in the sign of a AHC across the Fermi level is observed.}
    \label{fig:AHC_no_cap_Te}
\end{figure}
In the case of config.-I and -II, the surface of the MnTe films, due to uncompensated magnetization, gives rise to AHC at the Fermi level [see Fig. 3 in the main text]. Similar to the bulk MnTe, config.-III [Fig. \ref{fig:AHC_no_cap_Te}-(a)] with Te termination exhibit weak AHC at the Fermi level of $\sim$ 0.6$\mu S$. As discussed in the main text [see Fig. 3], the positive and negative $\Omega_{xy}^z$ are nearly equally compensated except in the vicinity of the high symmetry point $\Gamma$ to produce a weak AHC. On the contrary, for config-I and -II, there is a large uncompensated 
$\Omega_{xy}^z$ which gives rise to high AHC.
\begin{figure}
    \centering
    \includegraphics[width=0.7\linewidth]{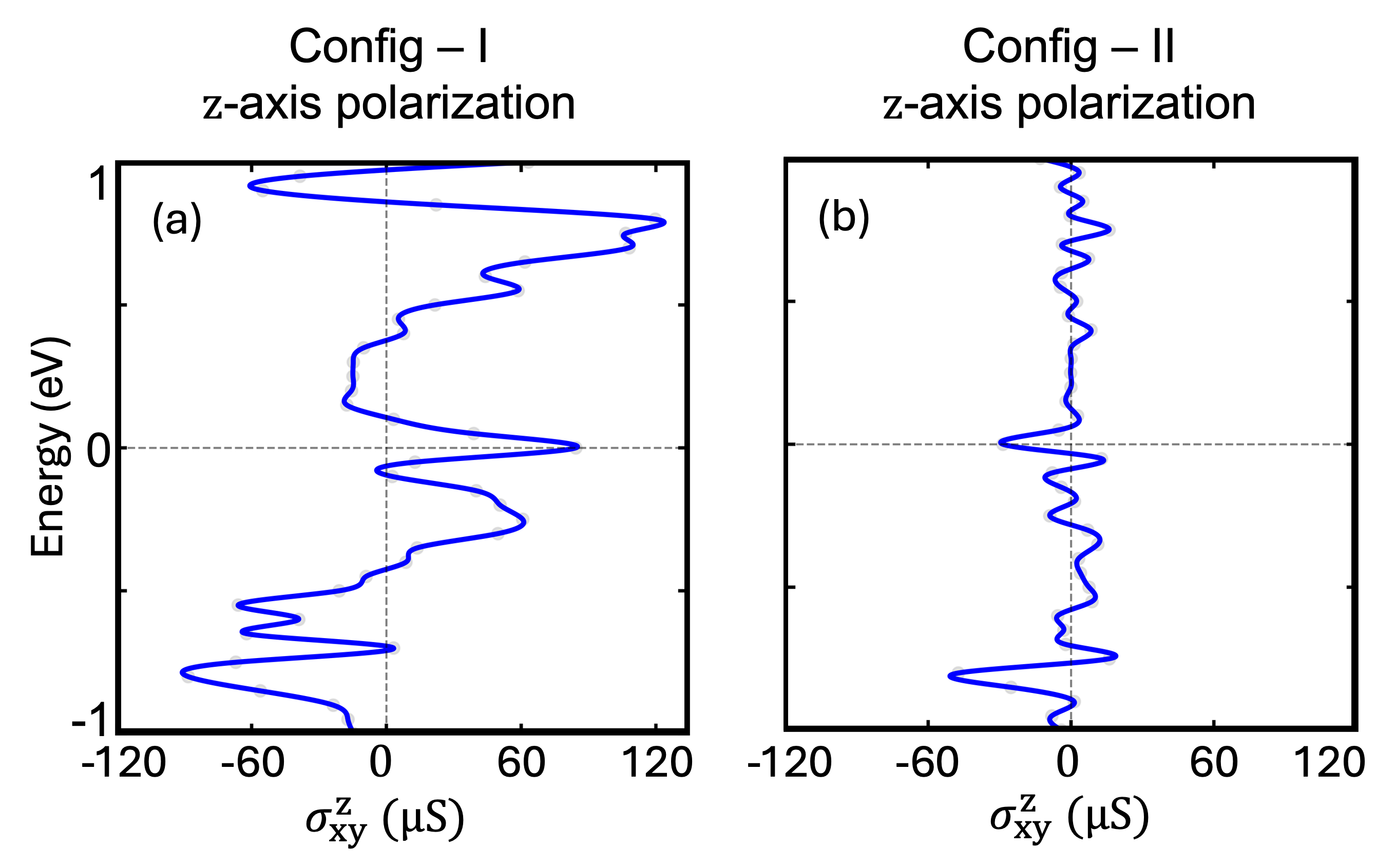}
    \caption{A finite AHC obtained for $z$ axis quantization for films as the six fold symmetry vanishes at the surface. Config-I (a) and config-II (b) both gives finite AHC at the Fermi.}
    \label{fig:AHC-z_quantization}
\end{figure}

As we know, the bulk MnTe belonging to $P6_{3}/mmc$ do not show AHC while spin quantization is along the $z$ axis. A symmetry analysis as presented in \cite{PhysRevB.111.184407} [see Table-II of the paper \cite{Das_2026}] shows that the existence of $\tau C_{6z}$ or $\tau S_{6z}$ prohibits the $\sigma_{xy}^z$ component in such cases. However, for the case of MnTe films as these symmetry vanishes the spin quantization along $z$ axis can give finite AHC at the Fermi [see Fig. \ref{fig:AHC-z_quantization}].

\subsection{Lifting of Krammers spin degeneracy through SOC}
\begin{figure} [htbp!]
    \centering
    \includegraphics[width=1.0\linewidth]{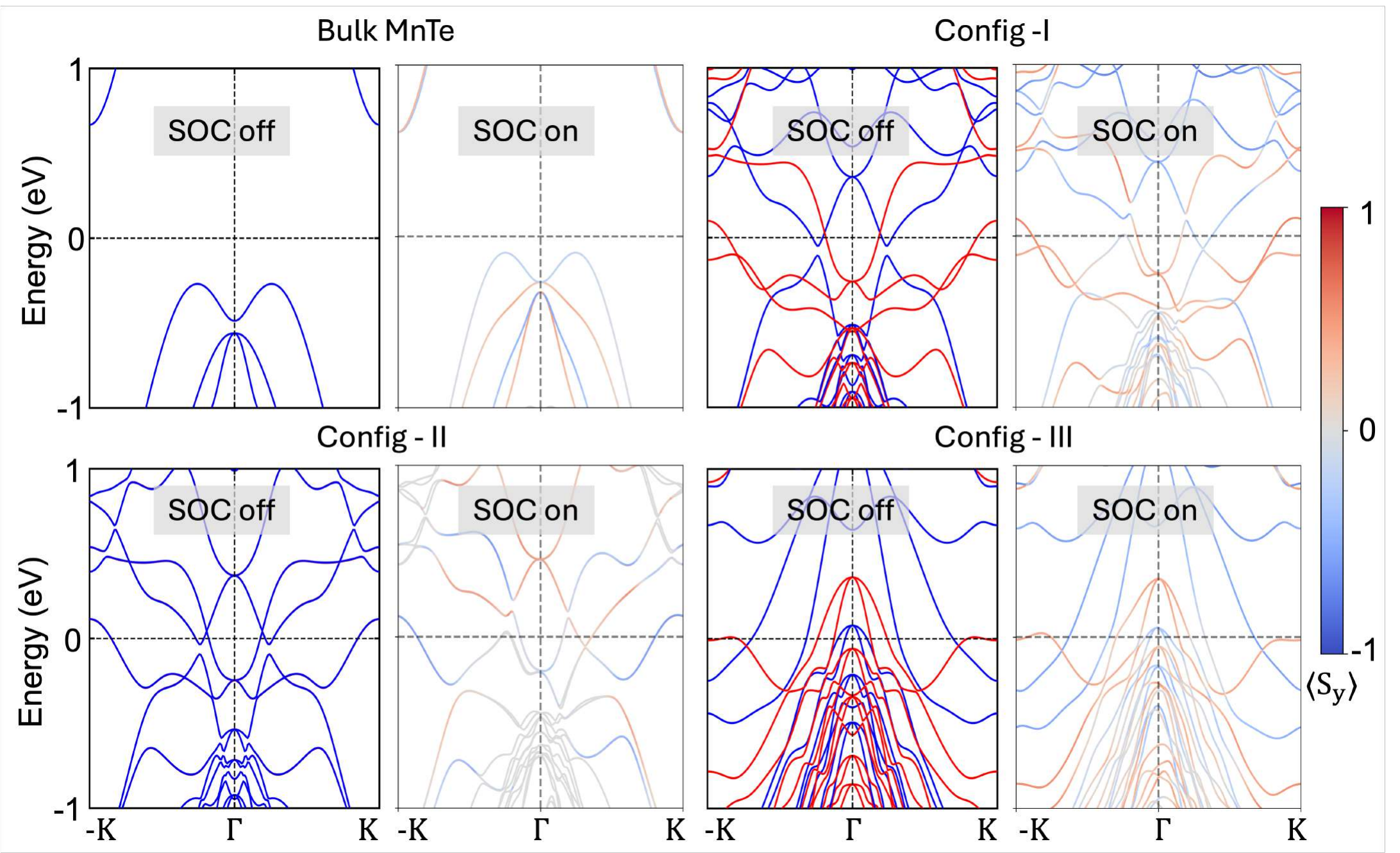}
    \caption{The band structure (with and without SOC) along the path $(-K)-\Gamma-K$ for MnTe bulk and films. The bulk MnTe and config-II [see Fig. 3 in the main text] exhibit Krammers spin degeneracy in the absence of SOC. The degeneracy is lifted in the presence of SOC.}
    \label{fig:LKSD}
\end{figure}
In Fig. \ref{fig:LKSD} the band dispersion for MnTe bulk and films are presented in the presence and absence of SOC. The lifting of Krammers spin degeneracy through SOC is intrinsically connected to the altermagnetic behavior of MnTe \cite{Krempasky_2024}.

\subsection{Effect of the structural relaxation on the bandstructure}
\begin{figure} [ht!]
    \centering
    \includegraphics[width=0.5\linewidth]{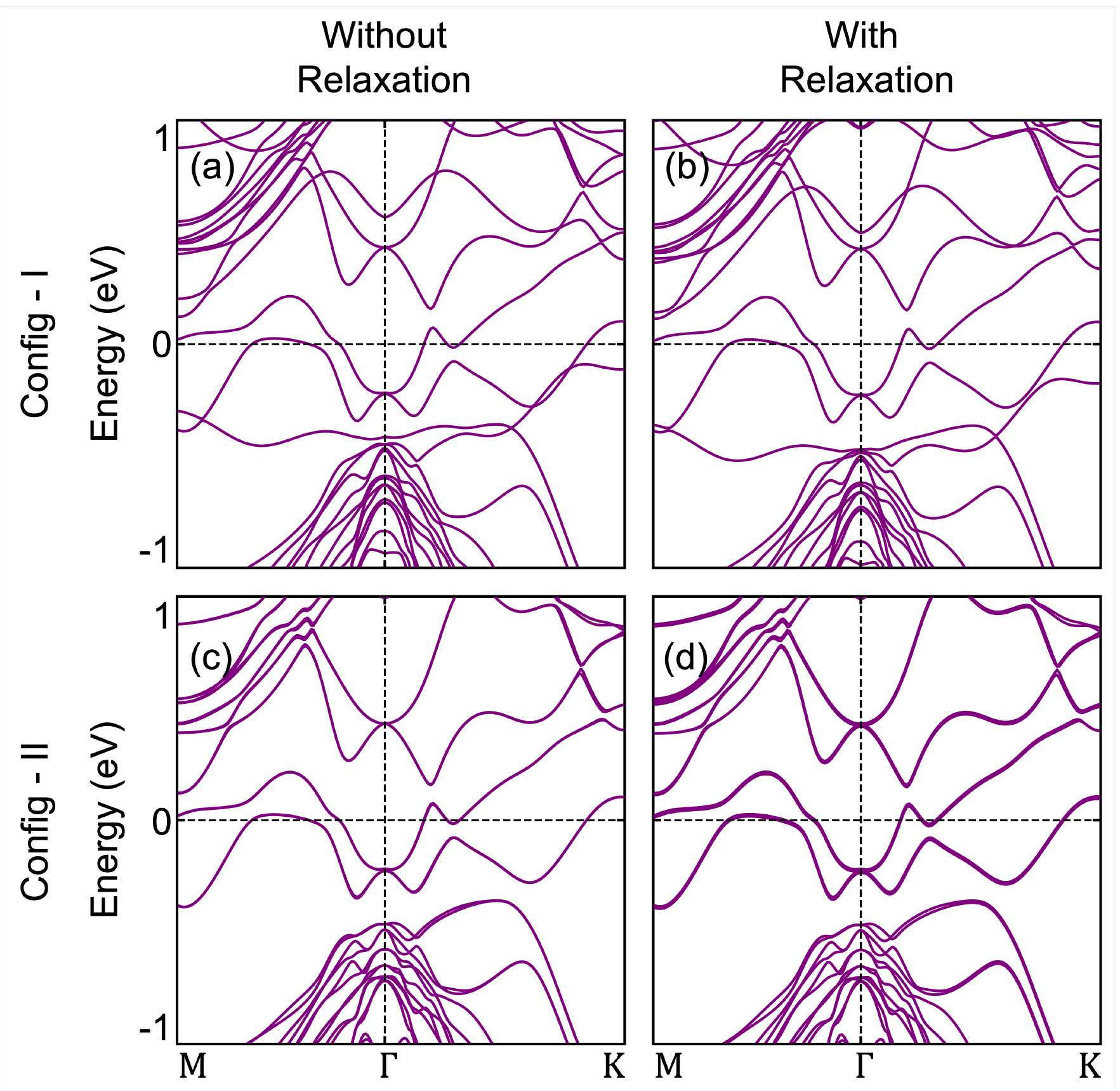}
    \caption{The DFT bandstructure for config-I on the top and config-II is at the bottom with their relaxed bandstructure on their right validates that relaxed bulk MnTe structure can be used for MnTe film studies.}
    \label{fig:DFT_band_comparison}
\end{figure}

The film designed from the experimental lattice parameters and after ionic relaxations does not show any discernible distortions and demonstrates negligible band dispersion [see Fig. \ref{fig:DFT_band_comparison}]. This suggests that the electronic structure remains largely invariant to the structural relaxation.

\subsection{Validation for the Wannier TB model}
\begin{figure} [htbp!]
    \centering
    \includegraphics[width=0.5\linewidth]{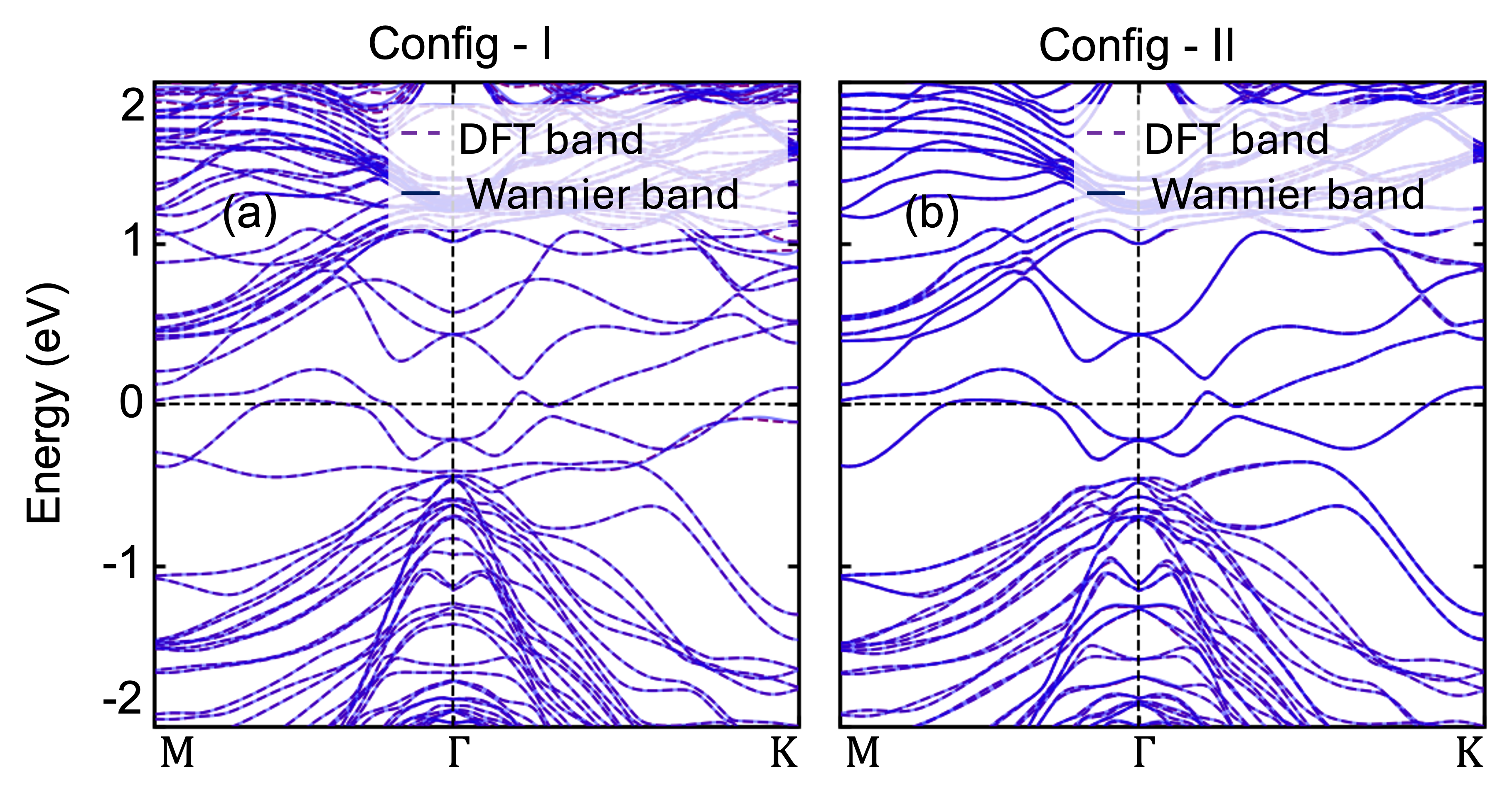}
    \caption{ The band structure obtained from DFT and Wannier TB model. The excellent agreement validates the accuracy of the maximally localized Wannier functions (MLWF) which are used to calculate the Berry curvature.}
    \label{fig:DFT_wannier}
\end{figure}
As can be seen from Fig. \ref{fig:DFT_wannier}, there is an excellent agreement between the band structures obtained from DFT and Wannier TB model. This validates the accuracy of MLWF  and therefore the Berry curvature and AHC.

\end{appendices}
\newpage

\bibliography{references.bib}